\documentclass[journal=jpccck,manuscript=letter,layout=traditional]{achemso}
\usepackage{chemformula} % Formula subscripts using \ch{}
\usepackage{booktabs} % Table markup
\usepackage{multirow} % Table rows spanning multiple cells
\usepackage{siunitx} % SI units
\usepackage{upgreek} % Non-italic Greek letters

\DeclareSIUnit\angstrom{\text{Å}}   % including Angstrom as a unit
\DeclareSIUnit\atm{\text{atm}}      % including atm as a unit
\DeclareSIUnit\amu{\text{u}}        % including amu as a unit

\newcommand{\twoone}{2_{1}}
\newcommand{\triplescrew}{$P\twoone\twoone\twoone$}
\newcommand{\centroscrew}{$P\twoone/c$}

\newcommand{\vectorchirality}{\boldsymbol{\upepsilon}}
\newcommand{\chirality}{\epsilon}

\newcommand{\hbasym}{\Delta r_{\mathrm{HB}}}
\newcommand{\epsA}{\epsilon_{\ch{A2}}}
\newcommand{\epsMX}{\epsilon_{\ch{MX4}}}
\newcommand{\epsMXpar}{\epsMX^{\parallel}}
\newcommand{\epsMXperp}{\epsMX^{\perp}}

\newcommand{\vecepsA}{\vectorchirality_{\ch{A2}}}
\newcommand{\vecepsMXpar}{\vectorchirality_{\ch{MX4}}^{\parallel}}
\newcommand{\vecepsMXperp}{\vectorchirality_{\ch{MX4}}^{\perp}}

\newcommand{\degreechi}{\chi^{x}}
\newcommand{\phiS}{\phi^{S}}
\newcommand{\phiR}{\phi^{R}}

\newcommand{\tauNH}{\tau_{\ch{NH3+}}}
\newcommand{\autocorrelationNH}{A_{\ch{NH3+}}}

\newcommand{\mba}{\ch{MBA2PbI4}} % Jana et al. Nat. Commun. (2020)
\newcommand{\smba}{(\textit{S}-\ch{MBA)2PbI4}} % Jana et al. Nat. Commun. (2020)
\newcommand{\rmba}{(\textit{R}-\ch{MBA)2PbI4}} % Jana et al. Nat. Commun. (2020)
\newcommand{\racmba}{(\textit{rac}-\ch{MBA)2PbI4}} % Dang et al. J. Phys. Chem. Lett. (2020)

\newcommand{\ronenea}{(\textit{R}-1\ch{NEA)2PbBr4}}

\newcommand{\rtwonea}{(\textit{R}-2\ch{NEA)2PbBr4}}

\newcommand{\rfourcl}{(\textit{R}-4-Cl-\ch{MBA)2PbBr4}}
\newcommand{\soneme}{(\textit{S}-1-Me-\ch{HA)2PbI4}}
\newcommand{\stwome}{(\textit{S}-2-Me-\ch{BuA)2PbBr4}}

\author{Mike Pols}
    \affiliation{Materials Simulation \& Modelling, Department of Applied Physics and Science Education, Eindhoven University of Technology, 5600 MB, Eindhoven, The Netherlands}
    \email{m.c.w.m.pols@tue.nl}
\author{Geert Brocks}
    \affiliation{Materials Simulation \& Modelling, Department of Applied Physics and Science Education, Eindhoven University of Technology, 5600 MB, Eindhoven, The Netherlands}
    \alsoaffiliation{Computational Chemical Physics, Faculty of Science and Technology and MESA+ Institute for Nanotechnology, University of Twente, 7500 AE, Enschede, The Netherlands}
    \email{g.h.l.a.brocks@tue.nl}
\author{Sof\'{i}a Calero}
    \affiliation{Materials Simulation \& Modelling, Department of Applied Physics and Science Education, Eindhoven University of Technology, 5600 MB, Eindhoven, The Netherlands}
    \email{s.calero@tue.nl}
\author{Shuxia Tao}
    \affiliation{Materials Simulation \& Modelling, Department of Applied Physics and Science Education, Eindhoven University of Technology, 5600 MB, Eindhoven, The Netherlands}
    \email{s.x.tao@tue.nl}

\title{Temperature-Dependent Chirality in Halide Perovskites}


\begin{document}

\begin{abstract}
With the use of chiral organic cations in two-dimensional metal halide perovskites, chirality can be induced in the metal halide layers, which results in semiconductors with intriguing chiral optical and spin-selective transport properties. The chiral properties strongly depend upon the temperature, despite the basic crystal symmetry not changing fundamentally. We identify a set of descriptors that characterize the chirality of metal halide perovskites such as \mba{}, and study their temperature dependence using molecular dynamics simulations with on-the-fly machine-learning force fields obtained from density functional theory calculations. We find that, whereas the arrangement of organic cations remains chiral upon increasing the temperature, the inorganic framework loses this property more rapidly. We ascribe this to the breaking of hydrogen bonds that link the organic with the inorganic substructures, which leads to a loss of chirality transfer.
\end{abstract}

In materials science, chirality refers to a structural asymmetry where two mirror images of a molecule or crystal structure can not be superimposed through any combination of translations and rotations. The mirror images of a crystal structure can commonly be connected via a mirror plane, inversion center or an improper rotation axis. The absence of such symmetry elements in a crystal signifies a chiral space group, for which the mirror images, or enantiomers, can be distinguished~\cite{nespoloCrystallographicShelvesSpacegroup2018}. A broad range of interesting properties emerge from chiral materials, including asymmetric chemical synthesis~\cite{feringaAbsoluteAsymmetricSynthesis1999}, circular dichroism (CD)~\cite{holzwarthUltravioletCircularDichroism1965} and circularly polarized photoluminescence (CPPL)~\cite{kumarCircularlyPolarizedLuminescence2015}, chirality-induced spin selectivity (CISS)~\cite{waldeckSpinSelectivityEffect2021}, particular forms of second-harmonic generation~\cite{sionckeSecondorderNonlinearOptical2003}, ferroelectricity~\cite{ruffChiralitydrivenFerroelectricityLiCuVO42019}, and topological quantum properties~\cite{changTopologicalQuantumProperties2018}. It has resulted in a wide interest in chiral materials over a broad range of domains in science.

With a focus on optoelectronic applications, chirality has been introduced in metal halide perovskites. Metal halide perovskites in particular exhibit excellent optoelectronic properties with a high tunability, making them promising candidates for solar cells~\cite{kimHighEfficiencyPerovskiteSolar2020}, light-emitting diodes (LEDs)~\cite{vanleRecentAdvancesHighEfficiency2018} and photodetectors~\cite{tianHybridOrganicInorganic2017}. The perovskite crystal lattice is a framework composed of vertex-linked inorganic \ch{MX6} octahedra (M = metal and X = halide), intercalated with inorganic or organic cations~\cite{akkermanWhatDefinesHalide2020}. Small cations result in the formation of three-dimensional (3D) \ch{MX6} frameworks, whereas larger organic cations facilitate the growth of lower dimensional frameworks; two-dimensional (2D), one-dimensional (1D), or zero-dimensional (0D)~\cite{etgarMeritPerovskiteDimensionality2018, linLowDimensionalOrganometalHalide2018}. The resulting materials, although 3D crystals, are commonly dubbed 2D, 1D, or 0D perovskites.

The flexibility of the perovskite crystal structure paves the way for the incorporation of chirality in halide perovskites by using large chiral organic ligands~\cite{billingBisVphenethylAmmonium2003, billingSynthesisCrystalStructures2006}. Lead and tin halide 2D perovskites currently attract most attention, and optical phenomena that are typical of chiral materials have been demonstrated in these materials, such as circular dichroism (CD)~\cite{ahnNewClassChiral2017, longSpinControlReduceddimensional2018, maChiral2DPerovskites2019, dangBulkChiralHalide2020, luHighlyDistortedChiral2020, ahnChiral2DOrganic2020, yangCircularlyPolarizedPhotoluminescence2022} and circularly polarized photoluminescence (CPPL)~\cite{longSpinControlReduceddimensional2018, maChiral2DPerovskites2019, dangBulkChiralHalide2020, yangCircularlyPolarizedPhotoluminescence2022, dursunTemperatureDependentOpticalStructural2023}. Concerning electronic transport, these materials also show a high degree of chirality-induced spin selectivity (CISS)~\cite{luSpindependentChargeTransport2019, luHighlyDistortedChiral2020}. The optoelectronic properties that draw most interest, typically stem from chirality present in the \ch{MX6} framework, as the electronic states around the band gap, which determine electronic transport and the onset of the optical transitions in these perovskites, originate from that framework. Since the latter does not inherently contain any chiral structural units, it requires a chiral structural distortion of this framework originating from the organic ligands.

By comparing various sets of chiral 2D perovskites, several studies have elucidated the relationship between structural chirality and the magnitude of spin splitting in electronic band structure~\cite{janaOrganictoinorganicStructuralChirality2020, janaStructuralDescriptorEnhanced2021}, as well as chiral optical activity~\cite{maElucidatingOriginChiroptical2022, sonUnravelingChiralityTransfer2023, apergiCalculatingCircularDichroism2023}. Although the fundamental mechanisms underlying this chiral optical activity and spin-selective transport are not well understood, these investigations have identified some distinct structural features that can be related to these processes. Specifically, a structural asymmetry in the tilting of the metal halide octahedra within the inorganic framework~\cite{janaOrganictoinorganicStructuralChirality2020, janaStructuralDescriptorEnhanced2021, apergiCalculatingCircularDichroism2023} and the hydrogen bonding at the interface between organic cations and the inorganic framework~\cite{maElucidatingOriginChiroptical2022, sonUnravelingChiralityTransfer2023} have been highlighted. Another significant experimental observation is the pronounced temperature dependence of the chiral optical properties of chiral perovskites~\cite{maChiral2DPerovskites2019, dursunTemperatureDependentOpticalStructural2023}, suggesting a decrease of chirality at elevated temperatures. To enhance the understanding of the temperature dependence of the chiral properties, it is essential to identify the chiral elements within the overall structure as well as the interplay between them and understand their evolution at finite temperatures.

In the present paper, we introduce a comprehensive set of structural descriptors that characterize chirality in 2D halide perovskite structures as a whole. Different descriptors capture the chirality in the arrangement of organic cations and the in-plane and out-of-plane chirality of the 2D \ch{MX6} framework, as illustrated for a selection of 2D lead halide perovskites. We study the temperature dependence of these structural chirality descriptors through molecular dynamics (MD) simulations with on-the-fly machine-learning force fields (MLFFs), based on density functional theory (DFT) calculations, which we apply to the archetype chiral 2D perovskite \mba{}. These simulations indicate that chirality in the organic cation arrangement persists up to a relatively high temperature, whereas the chirality in the lead halide planes is diminished substantially. The cause of this is found in the \ch{NH3+} groups that link the organic cations to the lead halide framework, whose rotations at elevated temperatures inhibit the chirality transfer from the cation to the framework.

\begin{figure*}[htbp]
    \includegraphics{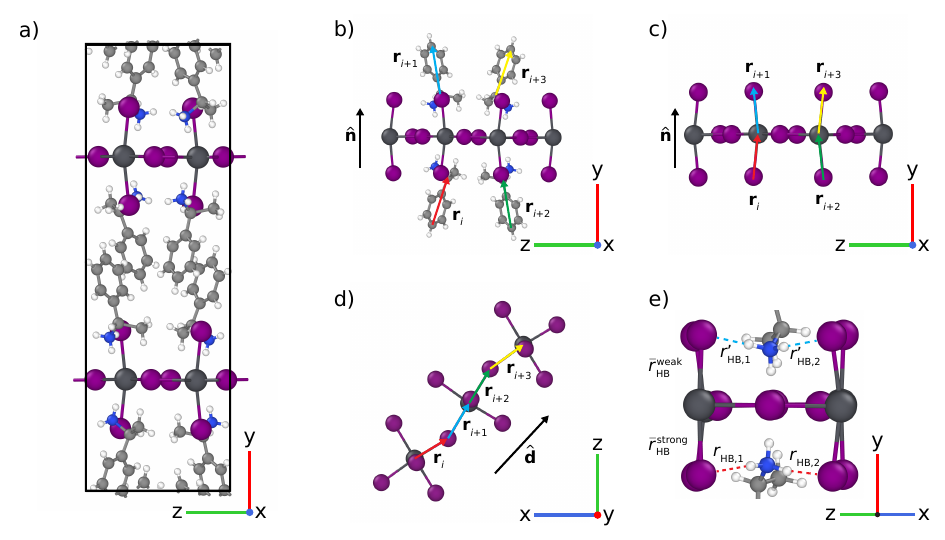}
    \caption{Overview of structural descriptors for two-dimensional (2D) metal halide perovskites. (a) Unit cell of chiral \smba{} perovskite. (b) Chirality of organic cations ($\epsA$). (c) Out-of-plane chirality of inorganic framework ($\epsMXperp$). (d) In-plane chirality of inorganic framework ($\epsMXpar$). (e) Hydrogen bond asymmetry ($\hbasym$). The various structural vectors used in the vector pairs and projection directions are shown in the figures.}
    \label{fig:structural_descriptors} 
\end{figure*}

To understand the transfer mechanism of structural chirality in halide perovskites, we begin by outlining the geometrical features we use to capture the symmetry breaking in the crystal structure of the archetype 2D chiral perovskite with its two enantiopure crystals (\textit{S}-\ch{MBA)2PbI4} and (\textit{R}-\ch{MBA)2PbI4}. Here, \textit{S}-/\textit{R}-\ch{MBA+} indicate the two enantiomers of the methylbenzylammonium cation (\ch{MBA+}). This crystal structure, which is shown in Figure~\ref{fig:structural_descriptors}a, is composed of layers of corner-sharing metal halide (\ch{PbI6}) octahedra, between which the \textit{S}-/\textit{R}-\ch{MBA+} cations are located. The compound crystallizes in the \triplescrew{} space group~\cite{janaOrganictoinorganicStructuralChirality2020}, which has a $2_{1}$ screw axis in the direction of all lattice vectors and lacks an inversion center, and is thus chiral. In contrast, a racemic mixture of \textit{S}-\ch{MBA+} and \textit{R}-\ch{MBA+} is found to crystallize in a \racmba{} structure with a \centroscrew{} space group~\cite{dangBulkChiralHalide2020}, which, besides its $2_{1}$ screw axis, also has an inversion center and is therefore achiral.

We employ a set of structural descriptors to probe the extent to which symmetries are broken in various parts of the crystal structures. The descriptors assess the chirality of the arrangement of the organic cations (Figure~\ref{fig:structural_descriptors}b), the out-of-plane and the in-plane chirality of the inorganic framework (Figure~\ref{fig:structural_descriptors}c and Figure~\ref{fig:structural_descriptors}d), as well as the asymmetry in the hydrogen bonding (Figure~\ref{fig:structural_descriptors}e). 
We find that these descriptors are more useful than others for characterizing structural chirality. Other possible descriptors, which measure distortions within and between the inorganic octahedra, are listed in Supporting Note 1.

In characterizing the chirality of various components of chiral perovskites, we use the concept of vector chirality. For magnetic systems, this concept has been used to assign a chirality or handedness to spin configurations~\cite{kawamuraSpinChiralityorderingsFrustrated2001, dingFieldtunableToroidalMoment2021, liAllelectricalReadingWriting2022}. Adapting the vector chirality concept from spin systems to crystal structures, we replace the spin orientations with the orientation vectors of chemical bonds. As such, we define the structural vector chirality $\vectorchirality$ as the sum over pairwise cross products of neighbouring orientation vectors as
\begin{equation}
    \label{eq:vector_chirality}
    \vectorchirality = \frac{1}{N} \sum^{N}_{i = 1} \hat{\mathbf{u}}_{i} \times \hat{\mathbf{u}}_{i + 1},
\end{equation}
where $\hat{\mathbf{u}}_{i} = \mathbf{r}_{i} / | \mathbf{r}_{i} |$ are the normalized bond orientation vectors and $N$ is the number of vectors within a selected set; see for example Figure~\ref{fig:structural_descriptors}b-d. To measure the handedness in a specific direction $\hat{\mathbf{p}}$, we project the vector chirality $\vectorchirality$ along that direction as
\begin{equation}
    \label{eq:projected_vector_chirality}
    \chirality = \vectorchirality \cdot \hat{\mathbf{p}}
\end{equation}
to obtain the scalar structural chirality $\chirality$. A nonzero value of $\chirality$ indicates a net handedness, with the sign making a distinction between left and right, whereas a value of zero implies a lack of chirality in that particular component and direction of the structure. 

We assign a structural vector chirality to individual parts of the structure, the first of which pertains to the arrangement of the organic cations. As a first step we assign a direction vector $\mathbf{r}_i$ to each organic cation, see Figure~\ref{fig:structural_descriptors}b. The exact definitions of these direction vectors for the various cations are shown in Supporting Note 2. Subsequently, we use the direction vectors of four cations along the inorganic layers (Figure~\ref{fig:structural_descriptors}b) to calculate the structural vector chirality according to Equation~\ref{eq:vector_chirality}. Note that cations marked by $\mathbf{r}_{i}$ and $\mathbf{r}_{i+3}$ are related by an action along an in-plane the $\twoone$ screw axis, as are the molecules marked by $\mathbf{r}_{i+1}$ and $\mathbf{r}_{i+2}$. The resulting chirality vector $\vecepsA$ is then projected according to Equation~\ref{eq:projected_vector_chirality}, in the direction normal to the metal halide plane $\epsA = \vecepsA \cdot \hat{\mathbf{n}}$.

To characterize the out-of-plane chirality of the inorganic framework, we use the $\ch{M}-\ch{X}$ bonds perpendicular to the inorganic layers to define direction vectors, and treat those in an analogous way to the direction vectors of the organic cations (Figure~\ref{fig:structural_descriptors}c). This gives an out-of-plane chirality vector $\vecepsMXperp$, which, projected normal to the inorganic planes, yields $\epsMXperp = \vecepsMXperp \cdot \hat{\mathbf{n}}$. The chirality vector of the inorganic framework in plane ($\vecepsMXpar$) is determined using the direction vectors of in-plane $\ch{M}-\ch{X}$ bonds, selected along lines $\hat{\mathbf{d}}$ that connect the inorganic octahedra, see Figure~\ref{fig:structural_descriptors}d. This chirality vector is subsequently projected along the direction of these lines as $\epsMXpar = \vecepsMXpar \cdot \hat{\mathbf{d}}$. Projection in the $\hat{\mathbf{d}}$-direction captures the in-plane chirality of the structure. Other possible projections are discussed in Supporting Note 4.

To make a connection with previous work on chirality in metal halide perovskites \cite{janaOrganictoinorganicStructuralChirality2020, maElucidatingOriginChiroptical2022, sonUnravelingChiralityTransfer2023}, we also probe the symmetry-breaking in the hydrogen bonds, which form the links between \ch{NH3} groups of the organic cations and halide ions in the inorganic framework. We quantify this asymmetry through the difference in the hydrogen bond lengths (i.e. $\ch{N}-\ch{H}\cdots\ch{X}$ bonds) between the two opposing cations above and below the inorganic layer (Figure~\ref{fig:structural_descriptors}e). For each cation we compute the average bond length of the two shortest hydrogen bonds $\bar{r}_{\mathrm{HB}}= ( r_{\mathrm{HB},1} + r_{\mathrm{HB},2} ) / 2$, where $r_{\mathrm{HB},1}$ and $r_{\mathrm{HB},2}$ refer to the shortest and second shortest hydrogen bonds. The hydrogen bond asymmetry $\hbasym$ is subsequently determined for every inorganic cage as $\hbasym = \bar{r}^{\mathrm{weak}}_{\mathrm{HB}} - \bar{r}^{\mathrm{strong}}_{\mathrm{HB}}$, with $\bar{r}^{\mathrm{weak}}_{\mathrm{HB}}$ and $\bar{r}^{\mathrm{strong}}_{\mathrm{HB}}$ the average hydrogen bond length of the cation with the longest and shortest hydrogen bonds, respectively. A nonzero value of $\hbasym$ implies a breaking of symmetry, whereas a value of zero indicates a highly symmetric hydrogen bonding of the cations above and below the metal halide plane.

Next, we use the above-mentioned structural descriptors to assess the structural chirality of a set of 2D perovskites from experiments. We optimized the structures using density functional theory (DFT) calculations with the SCAN exchange-correlation functional~\cite{sunStronglyConstrainedAppropriately2015} in VASP~\cite{kresseInitioMoleculardynamicsSimulation1994, kresseEfficiencyAbinitioTotal1996, kresseEfficientIterativeSchemes1996}. The full details of of this structural optimization can be found in Supporting Note 3. The resulting descriptor values are shown in Table~\ref{tab:structural_descriptors}. A more detailed overview of the structural descriptors in 2D perovskites is shown in Supporting Note 4.

Focusing on chiral \smba{}, we find nonzero values for the chirality of the arrangement of cations ($\epsA$ = +46.490$\times$10\textsuperscript{-3}) which through asymmetric hydrogen bonds ($\hbasym$ = \SI{0.036}{\angstrom}) transfer the breaking of this structural symmetry to the inorganic framework ($\epsMXpar$ = +4.534$\times$10\textsuperscript{-3} and $\epsMXperp$ = +1.560$\times$10\textsuperscript{-3}). Notably, we find that the structural descriptors appropriately distinguish between the structural enantiomers of \mba{}, as shown in Table S6 in Supporting Note 4. For example, chiral \rmba{}, the structural enantiomer of \smba{}, has identical descriptor values with an opposite sign, which indicates the opposite handedness of the two enantiomers. We note that we do not find an opposite sign for the hydrogen bond asymmetry, which stems from its definition that results in $\hbasym$ $\geq$ 0 in static structures. Moreover, for achiral \racmba{} the structural descriptors have values of zero ($\epsA = \epsMXperp = \epsMXpar = 0$ and $\hbasym$ = \SI{0.0}{\angstrom}), which implies an absence of any structural chirality in these perovskites.

Shifting our focus from \mba{} to the other chiral 2D perovskites, we are able to assess the variance in the structural chirality across perovskite structures. As a result of the molecule-dependent definition of the cation orientation vectors (Supporting Note 2), the chirality of the cation arrangements of widely different molecular species can not be directly compared. In contrast, due to the similarity of the inorganic framework across compounds, the structural chirality of the inorganic framework, both in-plane and out-of-plane, can be compared across compounds. The comparison shows that the various components of the perovskite lattice can exhibit a structural symmetry breaking to varying extents, independent of each other. For example, \ronenea{} has a high in-plane and out-of-plane framework chirality, whereas \soneme{} only has a high in-plane chirality with a rather small out-of-plane chirality. Interestingly, the comparison shows that a larger asymmetry in the hydrogen bonds does not necessarily imply a larger structural distortion in the inorganic framework. However, all chiral 2D perovskites exhibit a symmetry breaking in the various components of the crystal structure, as demonstrated by the nonzero structural descriptors. Finally, we do not observe a strong correlation between the structural chirality and spin-splitting in 2D perovskites (Supporting Note 5), indicating there are potentially more effects relevant to spin-splitting in addition to chiral symmetry breaking.

\begin{table*}[htbp]
    \caption{Structural descriptors probing structural chirality and bond asymmetry in chiral 2D perovskite structures.}
    \label{tab:structural_descriptors}
    \begin{tabular}{cccccc}
        \toprule
         Perovskite                     & Ref.                                                              & $\epsA$ ($\times$10\textsuperscript{-3})  & $\epsMXpar$ ($\times$10\textsuperscript{-3})  & $\epsMXperp$ ($\times$10\textsuperscript{-3}) & $\hbasym$ (\SI{}{\angstrom})  \\ \hline
        \smba{}                         & (\!\!~\citenum{janaOrganictoinorganicStructuralChirality2020})    & +46.490                                   & +4.534                                        & +1.560                                        & 0.036                         \\
        \ronenea{}                      & (\!\!~\citenum{janaOrganictoinorganicStructuralChirality2020})    & +4.481                                    & +12.629                                       & -9.319                                        & 0.058                         \\
        \rtwonea{}                      & (\!\!~\citenum{sonUnravelingChiralityTransfer2023})               & -10.791                                   & -4.394                                        & -2.021                                        & 0.107                         \\
        \rfourcl{}                      & (\!\!~\citenum{janaStructuralDescriptorEnhanced2021})             & -21.903                                   & +7.762                                        & +0.229                                        & 0.031                         \\
        \soneme{}                       & (\!\!~\citenum{janaStructuralDescriptorEnhanced2021})             & -13.333                                   & +13.230                                       & -0.648                                        & 0.033                         \\
        \stwome{}                       & (\!\!~\citenum{janaStructuralDescriptorEnhanced2021})             & +38.809                                   & -4.302                                        & +0.849                                        & 0.000                         \\ \bottomrule
    \end{tabular}
\end{table*}

Having defined descriptors capable of identifying structural chirality in perovskite structures, we now focus on analyzing the effects that finite temperatures have on the structural chirality. This analysis allows us to characterize the persistence of chirality as a function of temperature, gaining insights into the differences in chirality between the various components in the perovskite lattice, and the effectiveness of chirality transfer between those components. We will use the archetype 2D chiral perovskite, \mba{}, as an example.

To assess the effects of temperature, we use MLFFs to study the lattice dynamics of 2D perovskites with MD simulations. The MLFFs are trained against energies, forces and stresses from DFT calculations, using the on-the-fly learning scheme to sample structures from first-principles MD simulations as described in Refs.~\citenum{jinnouchiOntheflyMachineLearning2019, jinnouchiPhaseTransitionsHybrid2019}. The DFT reference data were obtained using the SCAN exchange-correlation functional in calculations in VASP~\cite{kresseInitioMoleculardynamicsSimulation1994, kresseEfficiencyAbinitioTotal1996, kresseEfficientIterativeSchemes1996}. The full details of the training procedure, the training sets and the validation of the trained MLFFs are shown in Supporting Note 6. We highlight that the trained MLFFs possess a high degree of transferability across perovskites structures. Specifically, the model trained on \smba{} describes both \rmba{} and \racmba{} with a high accuracy, as shown in Figure S5 and Figure S7.

\begin{figure*}[htbp]
    \includegraphics{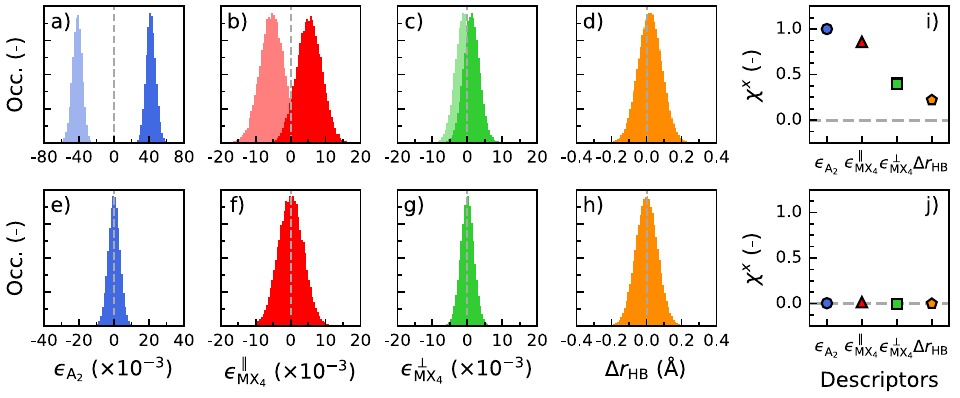}
    \caption{Finite-temperature distributions of structural descriptors and corresponding degree of chirality for \ch{MBA+}-based perovskites at \SI{50}{\K}. Distributions of (a) cation chirality, (b) in-plane framework chirality, (c) out-of-plane framework chirality, and (d) hydrogen bond asymmetry of \smba{} and \rmba{}. Distributions of (e) cation chirality, (f) in-plane framework chirality, (g) out-of-plane framework chirality, and (h) hydrogen bond asymmetry of \racmba{}. Degree of chirality for (i) \smba{} and (j) \racmba{}. In subfigures (a-d), the dark-colored distributions belong to \smba{}, whereas the light-colored distributions correspond to \rmba{}.}
    \label{fig:degree_of_chirality} 
\end{figure*}

Using the MLFFs, we subject the three variants of \ch{MBA+}-based perovskites, \smba{}, \rmba{}, and \racmba{}, to a temperature of \SI{50}{\K} at atmospheric pressure in $NpT$ MD simulations. As a result of thermal motion of the atoms within the structures the four descriptors ($\epsA$, $\epsMXpar$, $\epsMXperp$, and $\hbasym$) are no longer single-valued, but form distributions around mean values. The latter are close to the values reported for the static structures in Table~\ref{tab:structural_descriptors}, as shown in Figure~\ref{fig:degree_of_chirality}. Focusing on the descriptors that differ in sign between \smba{} and \rmba{} (Figure~\ref{fig:degree_of_chirality}a-c), we observe that in some cases the widths of the distributions due to thermal broadening are larger than the distance between the maxima of the peaks for \smba{} and \rmba{}. This is for example the case for the in-plane chirality ($\epsMXpar$; Figure~\ref{fig:degree_of_chirality}b) and, in particular, for the out-of-plane chirality ($\epsMXperp$; Figure~\ref{fig:degree_of_chirality}c) of the inorganic framework. Especially for $\epsMXperp$, thermal motion makes it very difficult to distinguish between \smba{} and \rmba{}. In contrast, the descriptor for the chirality in the organic cation arrangement, ($\epsA$; Figure~\ref{fig:degree_of_chirality}a), is only slightly affected by thermal motion. It can also be seen that the hydrogen bond asymmetry ($\hbasym$; Figure~\ref{fig:degree_of_chirality}d) is almost completely gone due to thermal motion. To support our findings, we also plot the descriptor distributions for \racmba{} (Figure~\ref{fig:degree_of_chirality}e-h), which are found to be symmetric around zero, indicating this crystal structure remains achiral at a finite temperature.

To quantify the extent to which the chirality has decreased as a result of temperature, we define a degree of chirality $\degreechi$ for a structural descriptor $x$ as
\begin{equation}
    \label{eq:degree_of_chirality}
    \degreechi = \frac{\phiS - \phiR}{\phiS + \phiR}
\end{equation}
with $\phiS$ and $\phiR$ representing the fractions of the distribution of $x$ which we associate with the \textit{S}- and \textit{R}-enantiomer, respectively. Per definition we have $-1 \leq \degreechi \leq +1$, with a value of $+1$ $(-1)$ corresponding to the ideal \textit{S}-form (\textit{R}-form) and a value of zero is associated with an achiral structure. Intermediate values indicate a diminished chirality with a preference for one of the forms. 

Figure~\ref{fig:degree_of_chirality}i shows the degrees of chirality for the different structural descriptors of \smba{}. Whereas the arrangement of cations ($\epsA$) has a the maximum degree of chirality $\degreechi$ = +1.00, the inorganic framework has considerably lower values with $\degreechi$ = +0.86 for the in-plane chirality ($\epsMXpar$), and +0.40 for the out-of-plane chirality ($\epsMXperp$). As noted above, the hydrogen bond asymmetry ($\hbasym$) suffers most from thermal motion, giving a small degree of chirality $\degreechi$ = +0.22. Again as a check, in \racmba{} we find a negligible degree of chirality for all descriptors ($\degreechi \approx 0$) as shown in Figure~\ref{fig:degree_of_chirality}j. 

A clear trend emerges from the results shown in Figure~\ref{fig:degree_of_chirality}. The chirality of the arrangement of organic cations is little affected by thermal motion at \SI{50}{\K}. In contrast, thermal vibrations in the inorganic framework limit the extent of the symmetry breaking and thus the degree of chirality in this framework. Concerning the latter, the axial halide anions, whose positions are used to determine $\epsMXperp$, are more mobile than the equatorial halide anions used to determine $\epsMXpar$, explaining the smaller degree of chirality of the former compared to the latter. The hydrogen bonding between organic cations and inorganic framework seems to be quite dynamic, resulting in a relatively small degree of chirality. 

\begin{figure*}[htbp]
    \includegraphics{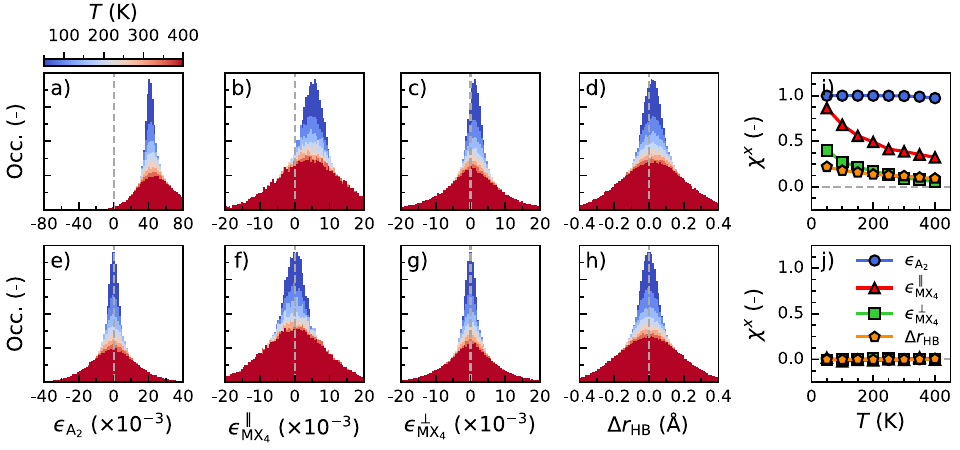}
    \caption{Temperature dependence of the degree of chirality in \ch{MBA+}-based perovskites. Temperature-dependent descriptor distributions for (a) cation chirality, (b) in-plane framework chirality, (c) out-of-plane framework chirality, and (d) hydrogen bond asymmetry in \smba{}. Temperature-dependent descriptor distributions for (e) cation chirality, (f) in-plane framework chirality, (g) out-of-plane framework chirality, and (h) hydrogen bond asymmetry in \racmba{}. Temperature-dependent degree of chirality for (i) \smba{} and (j) \racmba{}.}
    \label{fig:temperature_dependence_chirality} 
\end{figure*}

All of this indicates that the transfer of chirality from cations to inorganic framework is incomplete and suffers from thermal motion. This can be studied in more detail by focusing on the temperature dependence of the chiral descriptors. To do so, we increase the temperature from \SI{50}{\K} to \SI{400}{\K} in steps of \SI{50}{\K}, and show the resulting distributions and corresponding degrees of chirality in Figure~\ref{fig:temperature_dependence_chirality}. While the qualitative trends established above in Figure~\ref{fig:degree_of_chirality} remain, we observe an increase in the broadening of the distributions with increasing temperatures, because of increased thermal motion (Figure~\ref{fig:temperature_dependence_chirality}a-d). For chiral perovskites, it implies that the degree of chirality $\degreechi$ decreases with increasing temperature (Figure~\ref{fig:temperature_dependence_chirality}i). As a consistency check, we submit achiral \racmba{} to the same procedure. It shows a similar temperature broadening of the distributions (Figure~\ref{fig:temperature_dependence_chirality}e-h), but the degree of chirality for all descriptors remains zero, Figure~\ref{fig:temperature_dependence_chirality}j.

Interestingly, the degree of chirality decreases to a different extent for the various descriptors in \smba{}. The chirality of the arrangement of organic cations ($\epsA$) maintains a high degree of chirality for all investigated temperatures, only dropping to $\degreechi$ = +0.97 at \SI{400}{\K}. In contrast, the in-plane framework chirality ($\epsMXpar$) drops to a significantly lower degree of chirality of $\degreechi$ = +0.32, whereas the out-of-plane framework chirality ($\epsMXperp$) almost vanishes $\degreechi$ = +0.05 at \SI{400}{\K}. Likewise, the hydrogen bond asymmetry ($\hbasym$), which already had a low degree of chirality at low temperatures, drops to a small value of $\degreechi$ = +0.09.

Based on this, we conclude that the arrangement of cations is rather insensitive to temperature, whereas the in-plane and out-of-plane chirality of the inorganic framework are considerably more sensitive, exhibiting a major drop with increasing temperature. We attribute this to a decrease of chirality transfer from the organic cations to the inorganic framework with increasing temperature. A more detailed analysis of the temperature dependence, in particular its sensitivity to the type of structural enantiomer and the exchange-correlation functional the MLFF is trained against, is given in Supporting Note 7.

For a detailed understanding of the chirality transfer between the organic cations and the inorganic framework, we examine the interface between these two components. From our simulations we find, not surprisingly, that the organic cations are rigidly bonded frameworks with their atoms making relatively small vibrations around their equilibrium structure at finite temperature. The analysis above shows that even their arrangement and overall orientation changes little as a function of temperature. The main degrees of freedom within the organic cations, active at a relatively low temperature, are rotations around single bonds, where in the packed crystal structure rotations of the \ch{NH3+} head groups are the most accessible.

\begin{figure}[htbp]
    \includegraphics{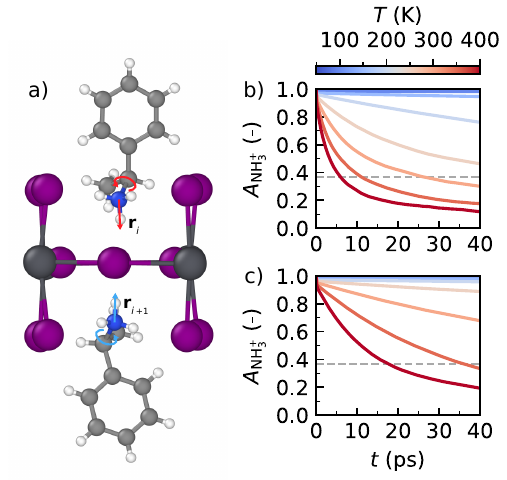}
    \caption{Orientational autocorrelation of \ch{NH3+} headgroup of organic cations. (a) Schematic overview of the $\ch{N}-\ch{H}$ bond vectors $\mathbf{r}_{i}$ used to determine the orientation of the headgroups of organic cations. Temporal autocorrelation of the headgroup orientation $A_{\ch{NH3+}}$ for the cations in (b) \smba{} and (c) \racmba{} at temperatures ranging from \SI{50}{\K} to \SI{400}{\K}. The dashed gray line indicates where $A_{\ch{NH3+}} = e^{-1}$.}
    \label{fig:chirality_transfer} 
\end{figure}

Therefore, we focus on the hydrogen bonds between the \ch{NH3+} groups of the organic cations and the neighbouring halide ions. Specifically, we assess the persistence of the $\ch{N}-\ch{H}\cdots\ch{X}$ hydrogen bonds as a function of increasing temperature, zooming in on the orientational autocorrelation function of the $\ch{N}-\ch{H}$ bonds (Figure~\ref{fig:chirality_transfer}a). This autocorrelation function $\autocorrelationNH$ is defined as
\begin{equation}
    \label{eqn:orientational_autocorrelation}
    \autocorrelationNH \left( t \right) = \left \langle \frac{1}{N} \sum^{N}_{i = 1} \mathbf{\hat{u}}_{i} \left( t_{0} \right) \cdot \mathbf{\hat{u}}_{i} \left( t_{0} + t \right) \right \rangle_{t_{0}},
\end{equation}
with $\mathbf{\hat{u}}_{i} = \mathbf{r}_{i} / | \mathbf{r}_{i} |$ the normalized orientation vector of the $i^{\mathrm{th}}$ $\ch{N}-\ch{H}$ bond, $N$ the number of vectors used in the analysis, $t$ the time delay and $t_{0}$ the time origin, where the autocorrelation function is averaged over different $t_0$. As the structures of the \ch{NH3+} groups are fairly rigid, a decay of $\autocorrelationNH$ is a clear sign of rotation of these groups. We define the orientation lifetime $\tauNH$ as the time at which the autocorrelation function falls off to a value of $\autocorrelationNH \left( \tauNH \right) = e^{-1}$.

Evaluating the dynamics of the cation headgroups in chiral \smba{} (Figure~\ref{fig:chirality_transfer}b), we find that $\autocorrelationNH$ is constant at low temperatures ($\leq$ \SI{100}{\K}), which indicates an absence of reorientations of the cation headgroups. For higher temperatures, we observe a decrease in $\autocorrelationNH$ and thus a reorientation of the $\ch{N}-\ch{H}$ bonds in the perovskite. The typical time for such reorientations is $\tauNH$ = \SI{6.0}{\ps} at \SI{400}{\K}, which indicates the hydrogen bonds between the organic cations and the inorganic framework lose their structure at relatively short timescales. This lack of persistence in the hydrogen bonding severely limits the ability of the cations to transfer their chirality to the inorganic framework, thus explaining the decreasing degree of chirality at elevated temperatures. We note that the autocorrelation functions do not decay to zero even after long times, which indicates a long-time order in the orientation of the organic cations. This order is present as a result of the absence of any large-scale reorientations of the organic cations, as demonstrated using the width and length orientation vectors of the cations in Supporting Note 8.

Similarly, one can monitor the autocorrelation functions in achiral \racmba{} (Figure~\ref{fig:chirality_transfer}c). A comparison to chiral \smba{} shows that the \ch{NH3+} headgroups in \racmba{} have a more persistent orientation, with headgroup reorientations only starting from \SI{200}{\K} and a typical reorientation time of $\tau_{\ch{NH3+}}$ = \SI{17.5}{\ps} for the achiral structure at \SI{400}{\K}. We propose the differences can be related to differences in the hydrogen bond strength in the two perovskites. The chiral perovskite structure is more distorted than the achiral structure and also has longer and thus weaker hydrogen bonds (Supporting Note 9), as is also discussed in Refs.~\citenum{janaOrganictoinorganicStructuralChirality2020, sonUnravelingChiralityTransfer2023}.

In summary, we introduce a set of structural descriptors to characterize the chirality in 2D halide perovskites, applicable to both static and finite temperature dynamical structures. The descriptors assess the chirality in the cations, the inorganic framework, and the asymmetry in the hydrogen bonds between the cations and inorganic layers. At finite temperatures, the structural descriptors no longer have a single value, but instead exhibit a spread around average values as a result of thermal motion. If the distributions for descriptors referring to opposite enantiomeric crystals overlap, the degree of chirality is reduced.

A comparison of the degree of chirality for the different structural descriptors reveals that the arrangement of the organic cations remains highly chiral up to a high temperature, whereas the chirality of the inorganic framework markedly decreases as a function of increasing temperature. Notably, a higher degree of chirality is observed for the in-plane direction than for the out-of-plane direction of the framework at elevated temperatures, which we explain through a larger mobility of the axial halide ions compared to the equatorial halide ions.

An analysis of the temperature dependence of the molecular motions reveals that, whereas the arrangement of organic cations is fairly stable with increasing temperature, the \ch{NH3+} headgroups can rotate. The reorientations of these groups undermine the rigidity of the hydrogen bonding between the organic cations and the inorganic framework, which weakens the transfer of chirality between cations and framework.

\begin{suppinfo}

\begin{itemize}
    \item Two-dimensional (2D) halide perovskite structures from experiments and density functional theory (DFT) calculations
    \item Additional structural descriptors for 2D halide perovskites; orientation vectors for cation chirality; computational settings for DFT calculations; additional details for structural descriptors; analysis of spin-splitting in 2D halide perovskites; training procedure and validation of machine-learning force fields (MLFFs); additional details for degree of chirality; additional details on chirality transfer; structural distortions and hydrogen bonds in perovskite structures
\end{itemize}

\end{suppinfo}

\begin{acknowledgement}
We thank Sofia Apergi for her help in the validation of the initial computation of the structural descriptors for chirality in a variety of chiral 2D metal halide perovskites. S.T. acknowledges funding from Vidi (project no. VI.Vidi.213.091) from the Dutch Research Council (NWO).
\end{acknowledgement}

\clearpage

%%% Bibliography
\bibliography{ms}

\end{document}

\title{Supporting Information: \\ Temperature-Dependent Chirality in Halide Perovskites}

\author{Mike Pols}
\email{m.c.w.m.pols@tue.nl}
\affiliation{%
    Materials Simulation \& Modelling, Department of Applied Physics and Science Education, Eindhoven University of Technology, 5600 MB, Eindhoven, The Netherlands
}%

\author{Geert Brocks}
\email{g.h.l.a.brocks@tue.nl}
\affiliation{%
    Materials Simulation \& Modelling, Department of Applied Physics  and Science Education, Eindhoven University of Technology, 5600 MB, Eindhoven, The Netherlands
}%
\affiliation{%
    Computational Chemical Physics, Faculty of Science and Technology and MESA+ Institute for Nanotechnology, University of Twente, 7500 AE, Enschede, The Netherlands
}%

\author{Sof\'{i}a Calero}
\email{s.calero@tue.nl}
\affiliation{%
    Materials Simulation \& Modelling, Department of Applied Physics and Science Education, Eindhoven University of Technology, 5600 MB, Eindhoven, The Netherlands
}%

\author{Shuxia Tao}
\email{s.x.tao@tue.nl}
\affiliation{%
    Materials Simulation \& Modelling, Department of Applied Physics and Science Education, Eindhoven University of Technology, 5600 MB, Eindhoven, The Netherlands
}%

\keywords{metal halide perovskites, density functional theory, machine-learning force fields, chirality, chiral perovskites, structural descriptors, symmetry breaking, structural chirality, bond asymmetry, chirality transfer}

\maketitle

\clearpage

\tableofcontents

\clearpage

\section{SI Note: Additional structural descriptor definitions}

\begin{figure*}[htbp]
    \includegraphics{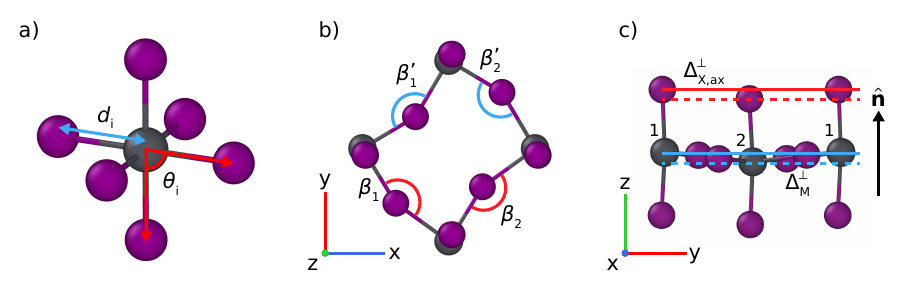}
    \caption{Additional structural descriptors used to characterize 2D perovskite structures. (a) Intraoctahedral distortions describing the variance in $\ch{M}-\ch{X}$ bond lengths and $\ch{X}-\ch{M}-\ch{X}$ bond angles. (b) Interoctahedral distortions that capture the cage asymmetry in the inorganic framework. (c) Planar distortions of the inorganic species away from the inorganic plane.}
    \label{fig:additional_structural_descriptors} 
\end{figure*}

In addition to the structural descriptors treated in the main text, we analyzed the two-dimensional (2D) perovskite structures with a range of additional descriptors that can be found in literature. The descriptors capture a variety of structural characteristics (Figure~\ref{fig:additional_structural_descriptors}), including distortions of the inorganic octahedra (intraoctahedral), the asymmetry of the inorganic cages in the layers of the inorganic framework (interoctahedral), and distortions of the inorganic species away from the inorganic lattice planes (planar). In the following sections it is detailed how each of these structural descriptors are computed. Results regarding these descriptors applied to 2D perovskites are shown in SI Note~\ref{subsection:additional_descriptors}.

\subsection{Intraoctahedral distortions} 

The distortions of the octahedra in the inorganic framework are quantified using metrics introduced by \citet{robinsonQuadraticElongationQuantitative1971} for the bond length and bond angle distortion. The bond length distortion $\Delta d$ is calculated as
\begin{equation}
    \label{eq:bond_length_distortion}
    \Delta d = \frac{1}{6} \sum^{6}_{i=1} \left( \frac{ d_{i} - d_{0}}{d_{0}} \right)^{2}
\end{equation}
with $d_{i}$ the bond length of one of the six $\ch{M}-\ch{X}$ bonds and $d_{0}$ the average length of all $\ch{M}-\ch{X}$ bonds in the \ch{MX6} octahedron. To determine the bond angle variance $\sigma^{2}$ we compute
\begin{equation}
    \label{eq:bond_angle_variance}
    \sigma^{2} = \frac{1}{11} \sum^{12}_{i=1} \left( \theta_{i} - \theta_{0} \right)^{2},
\end{equation}
with $\theta_{i}$ one of the twelve \textit{cis} $\ch{X}-\ch{M}-\ch{X}$ bond angles and $\theta_{0} = \SI{90}{\degree}$ the ideal value for such bond angles. An overview of the bonds and angles used in this computation is shown in Figure~\ref{fig:additional_structural_descriptors}a.

\subsection{Interoctahedral distortions} 

To probe the asymmetry of the inorganic cages we employ the disparity in adjacent angles of the inorganic cage as introduced by \citet{janaOrganictoinorganicStructuralChirality2020, janaStructuralDescriptorEnhanced2021}. The bond angle disparity $\deltabeta{}$ is computed as
\begin{equation}
    \label{eq:angle_disparity}
    \deltabeta{} = \beta - \beta'
\end{equation}
where $\beta$ and $\beta'$ ($\beta > \beta'$) are two unique $\ch{M}-\ch{X}-\ch{M}$ bond angles in a single inorganic cage. In static structures (Figure~\ref{fig:additional_structural_descriptors}b), the presence of two unique angles $\beta = \beta_{1} = \beta_{2}$ and $\beta' = \beta_{1}' = \beta_{2}'$ is enough to break the inversion symmetry, however during finite temperature simulations the thermal fluctuations distort the internal $\ch{M}-\ch{X}-\ch{M}$ angles, resulting in four unique angles: $\beta_{1}$, $\beta_{2}$, $\beta_{1}'$ and $\beta_{2}'$. To counteract these thermal fluctuations, we take the average value of the angle pairs
\begin{equation}
    \label{eq:average_angle}
    \begin{split}
        \Bar{\beta}  & = \left( \beta_{1}  + \beta_{2}  \right) / \, 2 \\
        \Bar{\beta}' & = \left( \beta_{1}' + \beta_{2}' \right) / \, 2
        \end{split}
\end{equation}
and consequently compute an averaged bond angle disparity using the following relation
\begin{equation}
    \label{eq:average_angle_disparity}
    \Delta \Bar{\beta} = \Bar{\beta} - \Bar{\beta}'.
\end{equation}
Although the above-mentioned bond angle disparity was computed for $\ch{M}-\ch{X}-\ch{M}$ angles in general, it should be noted that in-plane ($\deltabeta{in}$) and out-of-plane projections ($\deltabeta{out}$) can also be computed using the formulas derived by \citet{janaStructuralDescriptorEnhanced2021}.

Finally, using the values of the $\ch{M}-\ch{X}-\ch{M}$ angles in the inorganic cage, we can assign a value for the maximum distortion ($\maxD{}$) of the inorganic cage. This is done by selecting the smallest angle from the set of all angles $\{ \beta_{1}, \beta_{2}, \beta_{1}', \beta_{2}' \}$ and checking its deviation from $\SI{180}{\degree}$ as
\begin{equation}
    \label{eq:cage_distortion}
    \maxD{} = \SI{180}{\degree} - \min \left( \{ \beta_{1}, \beta_{2}, \beta_{1}', \beta_{2}' \} \right).
\end{equation}

\subsection{Planar distortions} 

To describe the breaking of symmetry in the inorganic layers, we employ planar distortion descriptors as introduced by \citet{apergiCalculatingCircularDichroism2023}. This symmetry breaking results from the existence of two unique octahedral sublattices in chiral halide perovskites (Figure~\ref{fig:additional_structural_descriptors}c). The planar distortion $\deltaperp$ is computed by making use of these two unique sublattices as
\begin{equation}
    \label{eq:planar_distortion}
    \Delta^{\perp} = \left( \mathbf{r}_{1} - \mathbf{r}_{2} \right) \cdot \hat{\mathbf{n}} = r^{\perp}_{1} - r^{\perp}_{2}
\end{equation}
where $\mathbf{r}_{1}$ and $\mathbf{r}_{2}$ are the position vectors of a species in octahedral sublattice 1 and 2, $\hat{\mathbf{n}}$ is the unit normal vector to the inorganic layer, and $r^{\perp}_{1}$ and $r^{\perp}_{2}$ are the position vectors projected on the unit normal vector. The planar distortion can be calculated for a variety of species, in this work we compute it for the metal ions \ch{M} ($\deltaMperp$) and axial halide species \ch{X}\textsubscript{ax} ($\deltaXaxperp$).

\clearpage

\section{SI Note: Cation orientation vectors}\label{section:atomic_fingerprints}

\subsection{Atomic fingerprints}

\begin{figure*}[htbp]
    \includegraphics{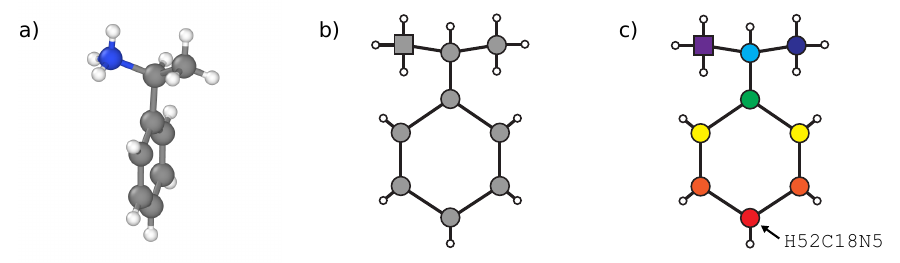}
    \caption{Demonstration of atomic fingerprinting in \ch{MBA+} cation. (a) Atomic structure of \ch{MBA+} cation (\ch{H} = white sphere; \ch{C} = gray sphere, and \ch{N} = blue sphere). (b) Graph representation of \ch{MBA+} cation in which the nodes represent atoms and the edges represent bonds (\ch{H} = white circle, \ch{C} = gray circle, and \ch{N} = gray square). (c) Graph representation of \ch{MBA+} cation with color coding of the \ch{C} and \ch{N} atoms according to their atomic fingerprint; \ch{C} and \ch{N} atoms with the same color have identical atomic fingerprints.}
    \label{fig:atomic_fingerprinting} 
\end{figure*}

To identify specific types of atoms in the topology of organic compounds (Figure~\ref{fig:atomic_fingerprinting}a), we assign a unique string to every atom in organic compounds through a process we refer to as fingerprinting. The identification string or atomic fingerprint is based on the local environment of each atom. Given an appropriate descriptor for the topology of the organic compound, all non-equivalent atoms in a compound can be uniquely identified, thus allowing for a straightforward method to identify atoms and orientation vectors in the molecules in the crystal structure.

Here, we assign atomic fingerprints using a graph representation of organic compounds (Figure~\ref{fig:atomic_fingerprinting}b), which we construct using the connectivity matrix of the atoms. The nodes of the graph are represented by the atoms themselves and the edges represent the bonds between the atoms. Using the molecular graph, a bond-based distance matrix of the graph is constructed, describing how many bonds are needed to connect two species with each other. To prevent expensive computations for large compounds, the computation of the bond-based distance matrix can be truncated up to a cutoff $N_{\mathrm{bonds}}$ beyond which species are not considered connected. Using the bond-based distance matrix, we compute the elemental distance sum for every atom $\mathbf{\upzeta}$ as
\begin{equation}
    \label{eq:elemental_distance_sum}
    \mathbf{\upzeta}_{i} = \mathbf{\upzeta}^{s}_{i} = \left( \zeta^{s_{1}}_{i}, \ldots, \zeta^{s_{N_{\mathrm{s}}}}_{i} \right) = \sum^{N_{\mathrm{bonds}}}_{j = 1} j \cdot N^{s}_{i,j}
\end{equation}
with $i$ iterating over the atoms in the system, $s$ iterating over the different species in the compounds, $j$ iterating over the various bond distances up to the cutoff $N_{\mathrm{bonds}}$. 
$N^{s}_{i,j}$ represents the number of atoms of species $s$ found at a bond distance $j$ from the $i$\textsuperscript{th} atom. The result of the procedure is that every atom is assigned a vector with $N_{s}$ dimensions (Figure~\ref{fig:atomic_fingerprinting}c). For ease-of-use these vectors were converted to a readable format by concatenating the elemental symbols with the elemental distance sums for all species as
\begin{equation}
    \label{eq:fingerprint_conversion}
    \texttt{H52C18N5} \leftarrow
    \begin{cases}
        \mathbf{\upzeta}_{i} & = \text{(52, 18, 5)} \\
        \mathbf{s}           & = \text{(H, C, N)}
    \end{cases}
\end{equation}
as is shown in Figure~\ref{fig:atomic_fingerprinting}c. Using this method with the \ch{MBA+} cation, a total of 13 unique atomic fingerprints were assigned in the cation; 1 for \ch{N}, 6 for \ch{C} and 6 for \ch{H}. The atomic fingerprints were subsequently used to extract a variety of internal vectors from the organic cations in the halide perovskites.

\subsection{Orientation vectors}

\begin{figure*}[htbp]
    \includegraphics{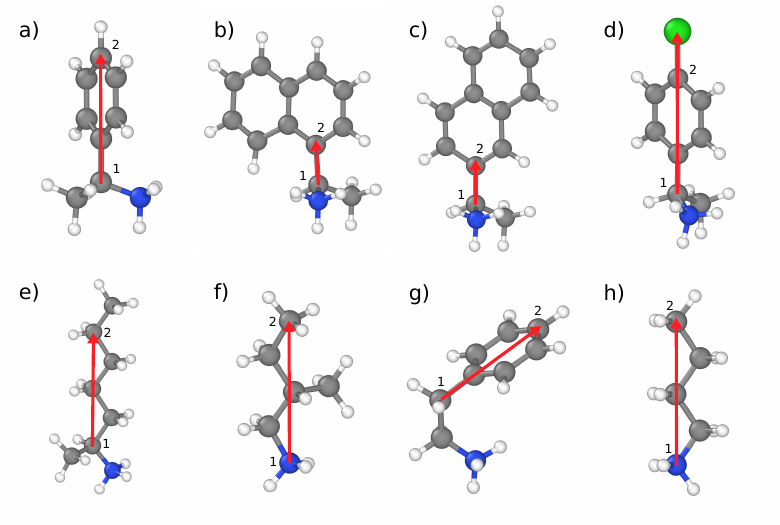}
    \caption{Orientation vectors for organic cations in 2D halide perovskites used to compute the chirality of the cation arrangement. The orientation vectors for the (a) \ch{MBA+}, (b) 1\ch{NEA+}, (c) 2\ch{NEA+}, (d) 4-\ch{Cl}-\ch{MBA+}, (e) 1-\ch{Me}-\ch{HA+}, (f) 2-\ch{Me}-\ch{BuA+}, (g) \ch{PEA+}, and (h) \ch{BA+}  cation are indicated with red arrows.}
    \label{fig:orientation_vectors_cations} 
\end{figure*}

An evaluation of the chirality of the arrangement of cations requires the definition of cation orientation vectors. With a wide range of atoms to choose from, it is important that the chosen orientation vectors are uniquely defined through the atomic fingerprints of the organic cation. The orientation vectors used in this work are shown in Figure~\ref{fig:orientation_vectors_cations} for which the atom fingerprints shown in Table~\ref{tab:orientation_vectors_fingerprints} were used to extract them from the perovskite crystal structures.

\begin{table}[htbp]
    \caption{Atom fingerprints used to determine the orientation vectors of various cations.}
    \label{tab:orientation_vectors_fingerprints} 
    \begin{ruledtabular}
    \begin{tabular}{ccccc}
        Cation                              & Element 1     & Atom fingerprint 1        & Element 2     & Atom fingerprint 2        \\ \hline
        \ch{MBA+}                           & \ch{C}        & \texttt{H32C16N1   }      & \ch{C}        & \texttt{H52C18N5   }      \\
        1\ch{NEA+}                          & \ch{C}        & \texttt{H45C32N1   }      & \ch{C}        & \texttt{H45C24N2   }      \\
        2\ch{NEA+}                          & \ch{C}        & \texttt{H47C36N1   }      & \ch{C}        & \texttt{H47C28N2   }      \\
        4-\ch{Cl}-\ch{MBA+}                 & \ch{C}        & \texttt{H27C16N1Cl5}      & \ch{Cl}       & \texttt{H62C26N6Cl0}      \\
        1-\ch{Me}-\ch{HA+}                  & \ch{C}        & \texttt{H59C16N1   }      & \ch{C}        & \texttt{H67C16N5   }      \\
        2-\ch{Me}-\ch{BuA+}                 & \ch{N}        & \texttt{H45C13N0   }      & \ch{C}        & \texttt{H45C9N4    }      \\
        \ch{PEA+}                           & \ch{C}        & \texttt{H34C16N2   }      & \ch{C}        & \texttt{H54C18N6   }      \\
        \ch{BA+}                            & \ch{N}        & \texttt{H36C10N0   }      & \ch{C}        & \texttt{H36C6N4    }      \\
    \end{tabular}
    \end{ruledtabular}
\end{table}

\clearpage

\section{SI Note: Density functional theory}\label{section:density_functional_theory}

To investigate the structural characteristics of the 2D perovskites, we carried out density functional theory (DFT) calculations for a variety of experimental crystal structures (Table~\ref{tab:perovskite_structures}). We performed these DFT calculations with the Vienna Ab-initio Simulation Package (VASP)~\cite{kresseInitioMoleculardynamicsSimulation1994, kresseEfficiencyAbinitioTotal1996, kresseEfficientIterativeSchemes1996} and modelled the electron-electron exchange interaction either using the SCAN~\cite{sunStronglyConstrainedAppropriately2015} or PBE+D3(BJ)~\cite{perdewGeneralizedGradientApproximation1996, grimmeEffectDampingFunction2011} functionals. The projector-augmented wave (PAW) pseudopotentials~\cite{kresseUltrasoftPseudopotentialsProjector1999} treated the following electrons: \ch{H} (1s\textsuperscript{1}), \ch{C} (2s\textsuperscript{2}2p\textsuperscript{2}), \ch{N} (2s\textsuperscript{2}2p\textsuperscript{3}), \ch{Cl} (3s\textsuperscript{2}3p\textsuperscript{5}), \ch{Br} (4s\textsuperscript{2}4p\textsuperscript{5}), \ch{I} (5s\textsuperscript{2}5p\textsuperscript{5}), and \ch{Pb} (6s\textsuperscript{2}6p\textsuperscript{2}) as valence electrons. The cutoff energy for the plane wave basis set was set to \SI{500}{\eV} in all calculations, combined with convergence criteria for the forces and energies of \SI{1E-5}{\eV} and \SI{1E-2}{\eV\per\angstrom}, respectively. The reciprocal space was sampled using a $\Gamma$-centered $k$-mesh~\cite{monkhorstSpecialPointsBrillouinzone1976}. In optimizing the experimental crystal structures all atomic positions, the cell shape and cell volume were allowed to change. The cell parameters of the experimental structures and the structures optimized using DFT calculations are shown in Table~\ref{tab:perovskites_optimized_1}, Table~\ref{tab:perovskites_optimized_2}, and Table~\ref{tab:perovskites_optimized_3}. We highlight that the SCAN functional performs better for the structures of the 2D perovskites than the PBE+D3(BJ) functional. SCAN predicts a unit cell volume of on average -1.5\% below that from experiments, whereas PBE+D3(BJ) underpredicts the unit cell volume with -3.2\%.

\begin{table}[htbp]
    \caption{Experimental crystal structures of investigated 2D halide perovskites with their space group and space group number, characterization temperature and database identification number.}
    \label{tab:perovskite_structures} 
    \begin{ruledtabular}
    \begin{tabular}{ccccc}
        Perovskite                  & Space group ($N_{\mathrm{group}}$)    & $T_{\mathrm{exp.}}$ (\SI{}{\K})   & Database ID                          & References                                                                    \\ \hline
        \smba{}                     & \triplescrew{} (19)                   & 298                               & \ccdc{2015617}                       & Ref.\citep{janaOrganictoinorganicStructuralChirality2020}                     \\
        \racmba{}                   & \centroscrew{} (14)                   & 293                               & \ccdc{1877052}                       & Ref.\citep{dangBulkChiralHalide2020}                                          \\ \hline
        \ronenea{}                  & \singlescrew{} (4)                    & 298                               & \ccdc{2015620}                       & Ref.\citep{janaOrganictoinorganicStructuralChirality2020}                     \\
        \sonenea{}                  & \singlescrew{} (4)                    & 298                               & \ccdc{2015618}                       & Ref.\citep{janaOrganictoinorganicStructuralChirality2020}                     \\
        \raconenea{}                & \centroscrew{} (14)                   & 298                               & \ccdc{2015614}                       & Ref.\citep{janaOrganictoinorganicStructuralChirality2020}                     \\ \hline
        \rtwonea{}                  & \singlescrew{} (4)                    & 298                               & \ccdc{2178604}                       & Ref.\citep{sonUnravelingChiralityTransfer2023}                                \\
        \stwonea{}                  & \singlescrew{} (4)                    & 298                               & \ccdc{2178605}                       & Ref.\citep{sonUnravelingChiralityTransfer2023}                                \\
        \ractwonea{}                & \centroscrew{} (14)                   & 298                               & \ccdc{2178606}                       & Ref.\citep{sonUnravelingChiralityTransfer2023}                                \\ \hline
        \rfourcl{}                  & \triplescrew{} (19)                   & 299.4                             & \ccdc{2095482}                       & Ref.\citep{janaStructuralDescriptorEnhanced2021}                              \\
        \soneme{}                   & \triplescrew{} (19)                   & 298.76                            & \ccdc{2095483}                       & Ref.\citep{janaStructuralDescriptorEnhanced2021}                              \\
        \stwome{}                   & \singlescrew{} (4)                    & 297.5                             & \ccdc{2095485}                       & Ref.\citep{janaStructuralDescriptorEnhanced2021}                              \\ \hline
        \pea{}                      & \ponebar{} (2)                        & 296                               & \ccdc{1542461}                       & Ref.\citep{duTwoDimensionalLeadII2017}                                        \\
        \ba{}                       & \pbca{} (61)                          & 100                               & \ccdc{2018893}                       & Ref.\citep{menahemStronglyAnharmonicOctahedral2021}                           \\
    \end{tabular}
    \end{ruledtabular}
\end{table}

\begin{sidewaystable}[htbp]
    \caption{Experimental and DFT-optimized crystal geometries for 2D halide perovskites.}
    \label{tab:perovskites_optimized_1} 
    \begin{ruledtabular}
    \begin{tabular}{cccccccccc}
        Perovskite                  & Type          & $a$ (\SI{}{\angstrom}) & $b$ (\SI{}{\angstrom}) & $c$ (\SI{}{\angstrom}) & $\alpha$ (\SI{}{\degree}) & $\beta$ (\SI{}{\degree}) & $\gamma$ (\SI{}{\degree}) & $V$ (\SI{}{\angstrom\cubed}) & $k$-points \\ \hline
        \multirow{3}{*}{\smba}      & Exp.          &  8.90 & 28.86 &  9.31 &  90.0 &  90.0 &  90.0 & 2393.31 & \multirow{3}{*}{\kpoints{2}{1}{2}} \\
                                    & SCAN          &  8.87 & 28.76 &  9.19 &  90.0 &  90.0 &  90.0 & 2342.49 & \\ 
                                    & PBE+D3(BJ)    &  8.76 & 28.53 &  9.12 &  90.0 &  90.0 &  90.0 & 2279.16 & \\ \hline
        \multirow{3}{*}{\racmba}    & Exp.          & 14.62 &  9.38 &  8.78 &  90.0 & 100.1 &  90.0 & 1185.67 & \multirow{3}{*}{\kpoints{2}{2}{2}} \\
                                    & SCAN          & 14.66 &  9.35 &  8.66 &  90.0 & 100.8 &  90.0 & 1166.97 & \\ 
                                    & PBE+D3(BJ)    & 14.56 &  9.28 &  8.55 &  90.0 &  99.9 &  90.0 & 1136.78 & \\ \hline
        \multirow{3}{*}{\pea}       & Exp.          &  8.74 &  8.74 & 33.00 &  84.6 &  84.7 &  89.6 & 2498.29 & \multirow{3}{*}{\kpoints{2}{2}{1}} \\
                                    & SCAN          &  8.70 &  8.72 & 32.76 &  85.6 &  85.7 &  89.4 & 2471.51 & \\ 
                                    & PBE+D3(BJ)    &  8.61 &  8.61 & 32.36 &  85.8 &  85.8 &  89.3 & 2386.81 & \\ \hline
        \multirow{3}{*}{\ba}        & Exp.          &  8.42 &  9.00 & 26.08 &  90.0 &  90.0 &  90.0 & 1975.59 & \multirow{3}{*}{\kpoints{2}{2}{1}} \\
                                    & SCAN          &  8.45 &  9.00 & 26.10 &  90.0 &  90.0 &  90.0 & 1986.58 & \\
                                    & PBE+D3(BJ)    &  8.34 &  8.96 & 26.00 &  90.0 &  90.0 &  90.0 & 1943.34 & \\
    \end{tabular}
    \end{ruledtabular}
\end{sidewaystable}

\clearpage

\begin{sidewaystable}[htbp]
    \caption{Experimental and DFT-optimized crystal geometries for 2D halide perovskites.}
    \label{tab:perovskites_optimized_2}  
    \begin{ruledtabular}
    \begin{tabular}{cccccccccc}
        Perovskite                  & Type          & $a$ (\SI{}{\angstrom}) & $b$ (\SI{}{\angstrom}) & $c$ (\SI{}{\angstrom}) & $\alpha$ (\SI{}{\degree}) & $\beta$ (\SI{}{\degree}) & $\gamma$ (\SI{}{\degree}) & $V$ (\SI{}{\angstrom\cubed}) & $k$-points \\ \hline
        \multirow{3}{*}{\ronenea}   & Exp.          &  8.76 &  7.96 & 19.52 &  90.0 &  93.8 &  90.0 & 1357.79 & \multirow{3}{*}{\kpoints{2}{2}{1}} \\
                                    & SCAN          &  8.67 &  7.91 & 19.40 &  90.0 &  94.4 &  90.0 & 1325.75 & \\ 
                                    & PBE+D3(BJ)    &  8.63 &  7.79 & 19.38 &  90.0 &  93.9 &  90.0 & 1299.72 & \\\hline
        \multirow{3}{*}{\sonenea}   & Exp.          &  8.75 &  7.96 & 19.50 &  90.0 &  93.8 &  90.0 & 1355.17 & \multirow{3}{*}{\kpoints{2}{2}{1}} \\
                                    & SCAN          &  8.67 &  7.91 & 19.42 &  90.0 &  94.5 &  90.0 & 1326.54 & \\ 
                                    & PBE+D3(BJ)    &  8.62 &  7.82 & 19.38 &  90.0 &  94.0 &  90.0 & 1303.01 & \\\hline
        \multirow{3}{*}{\raconenea} & Exp.          & 19.25 &  8.08 &  8.73 &  90.0 &  90.3 &  90.0 & 1357.21 & \multirow{3}{*}{\kpoints{1}{2}{2}} \\
                                    & SCAN          & 19.10 &  8.07 &  8.68 &  90.0 &  90.9 &  90.0 & 1337.41 & \\ 
                                    & PBE+D3(BJ)    & 19.14 &  7.96 &  8.63 &  90.0 &  90.9 &  90.0 & 1314.26 & \\\hline
        \multirow{3}{*}{\rtwonea}   & Exp.          &  8.77 &  7.84 & 20.35 &  90.0 &  99.5 &  90.0 & 1380.98 & \multirow{3}{*}{\kpoints{2}{2}{1}} \\
                                    & SCAN          &  8.70 &  7.78 & 20.33 &  90.0 &  99.7 &  90.0 & 1356.31 & \\ 
                                    & PBE+D3(BJ)    &  8.68 &  7.67 & 20.29 &  90.0 & 100.1 &  90.0 & 1329.32 & \\ \hline
        \multirow{3}{*}{\stwonea}   & Exp.          &  8.78 &  7.84 & 20.34 &  90.0 &  99.5 &  90.0 & 1380.50 & \multirow{3}{*}{\kpoints{2}{2}{1}} \\
                                    & SCAN          &  8.69 &  7.79 & 20.30 &  90.0 &  99.7 &  90.0 & 1354.53 & \\
                                    & PBE+D3(BJ)    &  8.68 &  7.67 & 20.27 &  90.0 & 100.1 &  90.0 & 1327.93 & \\ \hline
        \multirow{3}{*}{\ractwonea} & Exp.          & 20.13 &  7.91 &  8.74 &  90.0 &  92.5 &  90.0 & 1390.86 & \multirow{3}{*}{\kpoints{1}{2}{2}} \\
                                    & SCAN          & 19.98 &  7.80 &  8.69 &  90.0 &  92.3 &  90.0 & 1352.83 & \\
                                    & PBE+D3(BJ)    & 19.92 &  7.68 &  8.68 &  90.0 &  91.1 &  90.0 & 1329.32 & \\
    \end{tabular}
    \end{ruledtabular}
\end{sidewaystable}

\clearpage

\begin{sidewaystable}[htbp]
    \caption{Experimental and DFT-optimized crystal geometries for 2D halide perovskites.}
    \label{tab:perovskites_optimized_3} 
    \begin{ruledtabular}
    \begin{tabular}{cccccccccc}
        Perovskite                          & Type          & $a$ (\SI{}{\angstrom}) & $b$ (\SI{}{\angstrom}) & $c$ (\SI{}{\angstrom}) & $\alpha$ (\SI{}{\degree}) & $\beta$ (\SI{}{\degree}) & $\gamma$ (\SI{}{\degree}) & $V$ (\SI{}{\angstrom\cubed}) & $k$-mesh \\ \hline
        \multirow{3}{*}{\rfourcl}           & Experiment    &  7.91 &  8.81 & 35.58 &  90.0 &  90.0 &  90.0 & 2479.45 & \multirow{3}{*}{\kpoints{2}{2}{1}} \\
                                            & SCAN          &  7.86 &  8.71 & 35.60 &  90.0 &  90.0 &  90.0 & 2438.11 & \\
                                            & PBE+D3(BJ)    &  7.80 &  8.69 & 35.29 &  90.0 &  90.0 &  90.0 & 2390.75 & \\ \hline
        \multirow{3}{*}{\soneme}            & Experiment    &  8.97 &  8.97 & 33.78 &  90.0 &  90.0 &  90.0 & 2717.76 & \multirow{3}{*}{\kpoints{2}{2}{1}} \\
                                            & SCAN          &  8.88 &  8.88 & 33.93 &  90.0 &  90.0 &  90.0 & 2674.37 & \\
                                            & PBE+D3(BJ)    &  8.80 &  8.78 & 33.82 &  90.0 &  90.0 &  90.0 & 2615.19 & \\ \hline
        \multirow{3}{*}{\stwome}            & Experiment    & 15.69 &  8.29 &  8.24 &  90.0 & 101.2 &  90.0 & 1050.28 & \multirow{3}{*}{\kpoints{2}{2}{2}} \\
                                            & SCAN          & 15.29 &  8.21 &  8.18 &  90.0 & 101.4 &  90.0 & 1006.72 & \\
                                            & PBE+D3(BJ)    & 15.36 &  8.13 &  8.13 &  90.0 & 101.0 &  90.0 &  996.44 & \\
    \end{tabular}
    \end{ruledtabular}
\end{sidewaystable}

\clearpage

\section{SI Note: Structural descriptors in 2D perovskites}

\subsection{Chiral descriptors in 2D perovskites}

In addition to the chiral 2D perovskites shown in the main text, we also used the structural descriptors to evaluate the chirality in a variety of other perovskites. Here, we analyze the structural chirality of a variety of 2D perovskites found in previous works in literature shown in Table~\ref{tab:perovskite_structures}~\cite{janaOrganictoinorganicStructuralChirality2020, janaStructuralDescriptorEnhanced2021, dangBulkChiralHalide2020, menahemStronglyAnharmonicOctahedral2021, duTwoDimensionalLeadII2017, sonUnravelingChiralityTransfer2023}, which we optimized using DFT calculations. The results from this analysis can be found in Table~\ref{tab:structural_descriptors_1}, Table~\ref{tab:structural_descriptors_2}, and Table~\ref{tab:structural_descriptors_3}. The overview highlights the descriptors can distinguish between chiral and achiral perovskites as well as the handedness of the various components in the crystal structures of the 2D perovskites, as evidenced by the values and signs of the structural descriptors. We note that we mirrored the \smba{} structure to obtain the \rmba{} structure, resulting in exact mirror images. As a result of this, the values of the structural chirality descriptors are of equal magnitude but opposite sign. In contrast, independent crystal structures for both enantiomers were provided in the literature for \ch{NEA+}-based perovskites. These structures are not exactly identical mirror images, resulting in slight differences in the absolute values of the descriptors.

\begin{table}[htbp]
    \caption{Structural descriptors for chirality and bond asymmetry in 2D perovskite structures.}
    \label{tab:structural_descriptors_1} 
    \begin{ruledtabular}
    \begin{tabular}{ccccc}
        Perovskite      & $\epsA$  ($\times 10^{-3}$)   & $\epsMXpar$ ($\times 10^{-3}$)    & $\epsMXperp$ ($\times 10^{-3}$)   & $\hbasym$  (\SI{}{\angstrom})     \\ \hline
        \smba{}         & +46.490                       & +4.534                            & +1.560                            & 0.036                             \\
        \rmba{}         & -46.490                       & -4.534                            & -1.560                            & 0.036                             \\
        \racmba{}       & 0.000                         & 0.000                             & 0.000                             & 0.000                             \\ \hline
        \pea{}          & 0.000                         & 0.000                             & 0.000                             & 0.000                             \\
        \ba{}           & 0.000                         & 0.000                             & 0.000                             & 0.000                             \\
    \end{tabular}
    \end{ruledtabular}
\end{table}

\begin{table}[htbp]
    \caption{Structural descriptors for chirality and bond asymmetry in 2D perovskite structures.}
    \label{tab:structural_descriptors_2} 
    \begin{ruledtabular}
    \begin{tabular}{ccccc}
        Perovskite      & $\epsA$  ($\times 10^{-3}$)   & $\epsMXpar$ ($\times 10^{-3}$)    & $\epsMXperp$ ($\times 10^{-3}$)   & $\hbasym$  (\SI{}{\angstrom})     \\ \hline
        \sonenea{}      & -4.341                        & -13.105                           & +9.172                            & 0.044                             \\
        \ronenea{}      & +4.481                        & +12.629                           & -9.319                            & 0.058                             \\
        \raconenea{}    & 0.000                         & 0.000                             & 0.000                             & 0.000                             \\ \hline
        \stwonea{}      & +10.701                       & +5.519                            & +1.787                            & 0.098                             \\
        \rtwonea{}      & -10.791                       & -4.394                            & -2.021                            & 0.107                             \\
        \ractwonea{}    & 0.000                         & 0.000                             & 0.000                             & 0.000                             \\
    \end{tabular}
    \end{ruledtabular}
\end{table}

\begin{table}[htbp]
    \caption{Structural descriptors for chirality and bond asymmetry in 2D perovskite structures.}
    \label{tab:structural_descriptors_3} 
    \begin{ruledtabular}
    \begin{tabular}{ccccc}
        Perovskite      & $\epsA$  ($\times 10^{-3}$)   & $\epsMXpar$ ($\times 10^{-3}$)    & $\epsMXperp$ ($\times 10^{-3}$)   & $\hbasym$  (\SI{}{\angstrom})     \\ \hline
        \rfourcl{}      & -21.903                       & +7.762                            & +0.229                            & 0.031                             \\
        \soneme{}       & -13.333                       & +13.230                           & -0.648                            & 0.033                             \\
        \stwome{}       & +38.809                       & -4.302                            & +0.849                            & 0.000                             \\
    \end{tabular}
    \end{ruledtabular}
\end{table}

\clearpage

\subsection{Components of in-plane framework chirality}

\begin{figure*}[htbp]
    \includegraphics{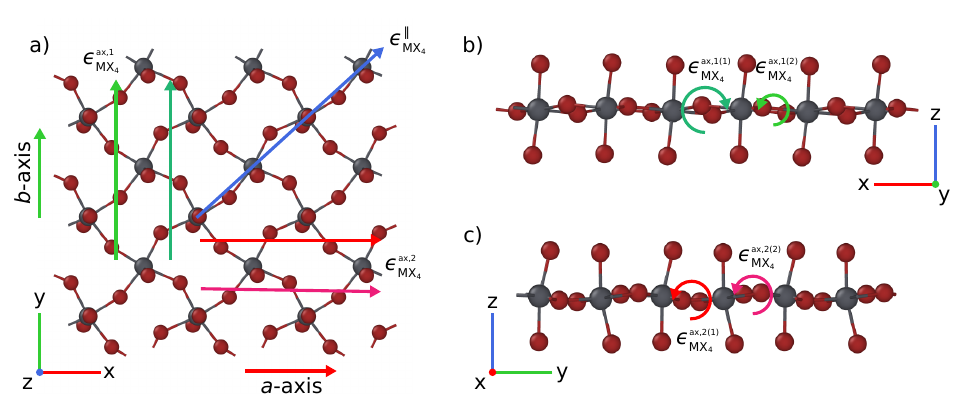}
    \caption{Decomposition of in-plane framework chirality into components along axial directions. (a) Overview of the various directions and framework chirality considered in \sonenea{}. (b) Two unique chirality values are found in the direction of the $b$-axis that is parallel to the $2_{1}$ screw axis and (c) only a single value in the direction of the $a$-axis.}
    \label{fig:in_plane_components} 
\end{figure*}

As indicated in the manuscript, the in-plane framework chirality ($\epsMXpar$) can be decomposed into various components. By determining the chirality in the direction of the axes spanning the inorganic layers, we obtain chirality components that probe the structural chirality along those directions (Table~\ref{tab:structural_chirality_components}). An example of such a decomposition of the in-plane framework chirality is shown for \sonenea{} in Figure~\ref{fig:in_plane_components}. The net chirality in the axial directions ($\epsMXaxone$ and $\epsMXaxtwo$) is obtained by averaging the various unique chirality values in those directions. For \sonenea{} we find two unique values for the framework chirality in the direction parallel to the $2_{1}$ screw axis ($b$-axis). The two values have an opposite sign and magnitude ($\epsMXaxoneone$ = -143.524$\times$10\textsuperscript{-3} and $\epsMXaxonetwo$ = +71.620$\times$10\textsuperscript{-3}), thus resulting in a net nonzero chirality in this direction of $\epsMXaxone$ = -35.952$\times$10\textsuperscript{-3} in the direction of the $b$-axis. Along the other axis, the $a$-axis, we find only a single value for the chirality ($\epsMXaxtwoone = \epsMXaxtwotwo$ = +14.691$\times$10\textsuperscript{-3}) which is thus also the net chirality of the inorganic layers in this direction, $\epsMXaxtwo$ = +14.691$\times$10\textsuperscript{-3}. We find the sum of the net chirality in the two axial directions roughly approximates the in-plane framework chirality for all investigated perovskites ($\epsMXaxone + \epsMXaxtwo \approx \epsMXpar$), thus indicating the in-plane framework chirality appropriately probes the chirality in all directions of the inorganic layers. Finally, we note that in achiral perovskites, such as \raconenea{} and \ractwonea{}, one does find nonzero chirality values in the inorganic layers. However, due to an opposite sign and identical magnitude, there is a nonzero net chirality in the axial directions in the inorganic layers.

\begin{sidewaystable}[htbp]
    \caption{Components of the in-plane framework chirality in 2D perovskite structures.}
    \label{tab:structural_chirality_components} 
    \begin{ruledtabular}
    \begin{tabular}{cccccccc}
        \multirow{3}{*}{Perovskite}     &                           & \multicolumn{3}{c}{Direction 1$^{*}$}                                     & \multicolumn{3}{c}{Direction 2$^{*}$}                                     \\
                                        & $\epsMXpar$               & $\epsMXaxone$             & $\epsMXaxoneone$      & $\epsMXaxonetwo$      & $\epsMXaxtwo$             & $\epsMXaxtwoone$      & $\epsMXaxtwotwo$      \\ 
                                        & ($\times 10^{-3}$)        & ($\times 10^{-3}$)        & ($\times 10^{-3}$)    & ($\times 10^{-3}$)    & ($\times 10^{-3}$)        & ($\times 10^{-3}$)    & ($\times 10^{-3}$)    \\ \hline
        \sonenea{}                      & -13.105                   & -35.952                   & -143.524              & +71.620               & +14.691                   & +14.691               & +14.691               \\ 
        \ronenea{}                      & +12.629                   & +40.313                   & +150.501              & -69.876               & -19.375                   & -19.375               & -19.375               \\ 
        \raconenea{}                    & 0.000                     & 0.000                     & 0.000                 & 0.000                 & 0.000                     & $+$12.727             & $-$12.727             \\ \hline
        \stwonea{}                      & +5.519                    & +30.709                   & +143.512              & -82.094               & -21.166                   & -21.166               & -21.166               \\ 
        \rtwonea{}                      & -4.394                    & -34.560                   & -147.671              & +78.550               & +26.102                   & +26.102               & +26.102               \\ 
        \ractwonea{}                    & 0.000                     & 0.000                     & 0.000                 & 0.000                 & 0.000                     & $+$30.614             & $-$30.614             \\ 
    \end{tabular}
    \end{ruledtabular}
\end{sidewaystable}

\clearpage

\subsection{Effects of geometry optimization on structural descriptors}

In addition to the computation of the structural descriptors for structures optimized with the SCAN functional, we also determined the values of the structural descriptors for the experimental structure and the PBE+D3(BJ)-optimized structures of \ch{MBA2PbI4}. The results of this analysis are shown in Table~\ref{tab:optimization_effects}. The nature of the crystal structures appears to have only a small effect on the values of the structural descriptors, indicating that the descriptors are rather robust in identifying structural chirality in 2D perovskites.

\begin{table}[htbp]
    \caption{Effect of structural optimizations on structural chirality in \ch{I}-based 2D perovskites.}
    \label{tab:optimization_effects} 
    \begin{ruledtabular}
    \begin{tabular}{cccccc}
        \multirow{2}{*}{Perovskite} & \multirow{2}{*}{Type} & $\epsA$               & $\epsMXpar$           & $\epsMXperp$          & $\hbasym$             \\ 
                                    &                       & ($\times 10^{-3}$)    & ($\times 10^{-3}$)    & ($\times 10^{-3}$)    & (\SI{}{\angstrom})    \\ \hline
        \multirow{3}{*}{\smba}      & Exp.                  & +49.313               & +5.252                & +1.341                & 0.029                 \\ 
                                    & SCAN                  & +46.490               & +4.534                & +1.560                & 0.036                 \\ 
                                    & PBE+D3(BJ)            & +44.446               & +5.917                & +1.578                & 0.031                 \\ \hline
        \multirow{3}{*}{\rmba}      & Exp.                  & -49.313               & -5.252                & -1.341                & 0.029                 \\ 
                                    & SCAN                  & -46.490               & -4.534                & -1.560                & 0.036                 \\ 
                                    & PBE+D3(BJ)            & -44.446               & -5.917                & -1.578                & 0.031                 \\ \hline
        \multirow{3}{*}{\racmba}    & Exp.                  & 0.000                 & 0.000                 & 0.000                 & 0.000                 \\ 
                                    & SCAN                  & 0.000                 & 0.000                 & 0.000                 & 0.000                 \\ 
                                    & PBE+D3(BJ)            & 0.000                 & 0.000                 & 0.000                 & 0.000                 \\
    \end{tabular}
    \end{ruledtabular}
\end{table}

\clearpage

\begin{sidewaystable}[htbp]
    \subsection{Additional structural descriptors}\label{subsection:additional_descriptors}
    \caption{Additional structural descriptors as determined for crystal geometries optimized with the SCAN functional.}
    \label{tab:additional_structural_descriptors} 
    \begin{ruledtabular}
    \begin{tabular}{cccccccccccc}
        \multirow{2}{*}{Perovskite} & $\sigmatwo$    & $\deltad$          & $\deltabeta{}$ & $\deltabeta{in}$ & $\deltabeta{out}$ & $\maxD{}$       & $\maxD{in}$     & $\maxD{out}$    & $\deltaMperp$        & $\deltaXaxperp$      \\
                                    & ($^{\circ 2}$) & ($\times 10^{-5}$) & ($^{\circ}$)   & ($^{\circ}$)     & ($^{\circ}$)      & ($^{\circ}$)    & ($^{\circ}$)    & ($^{\circ}$)    & (\SI{}{\angstrom{}}) & (\SI{}{\angstrom{}}) \\ \hline
        \smba{}                     & 20.17 & 29.38  &  7.383 &  7.376 &  0.652 & 31.352 & 31.345 &  0.748 & 0.061 & 0.043 \\ 
        \rmba{}                     & 20.17 & 29.38  &  7.383 &  7.376 &  0.652 & 31.352 & 31.345 &  0.748 & 0.061 & 0.043 \\ 
        \racmba{}                   & 14.04 & 18.19  &  0.000 &  0.000 &  0.000 & 26.193 & 26.191 &  0.293 & 0.000 & 0.000 \\ \hline
        \pea{}                      &  4.10 &   4.47 &  1.622 &  1.649 & -1.180 & 31.476 & 31.476 &  1.623 & 0.000 & 0.009 \\ 
        \ba{}                       &  7.29 &   0.54 &  0.000 &  0.000 &  0.000 & 32.933 & 23.964 & 23.250 & 0.000 & 0.000 \\ \hline
        \sonenea{}                  & 42.48 &  75.48 & 16.075 & 18.103 & -3.781 & 39.957 & 39.402 & 11.113 & 0.142 & 0.126 \\ 
        \ronenea{}                  & 43.21 &  70.85 & 15.980 & 18.324 & -4.554 & 39.968 & 39.443 & 11.688 & 0.135 & 0.111 \\ 
        \raconenea{}                & 22.58 &  20.88 &  0.000 &  0.000 &  0.000 & 27.844 & 27.826 &  1.056 & 0.000 & 0.000 \\ \hline
        \stwonea{}                  & 48.15 &  47.24 & 15.997 & 15.341 &  5.229 & 40.092 & 38.506 & 12.038 & 0.072 & 0.182 \\ 
        \rtwonea{}                  & 49.53 &  48.14 & 15.946 & 15.078 &  6.024 & 40.555 & 38.863 & 12.517 & 0.058 & 0.170 \\ 
        \ractwonea{}                & 23.43 &   8.97 &  0.000 &  0.000 &  0.000 & 31.970 & 31.878 &  2.550 & 0.000 & 0.000 \\ \hline
        \rfourcl{}                  & 52.46 &  72.28 & 14.266 & 10.782 & 14.690 & 39.383 & 35.729 & 17.771 & 0.113 & 0.201 \\ 
        \soneme{}                   & 10.91 &   3.33 &  1.033 &  1.026 &  0.163 & 29.372 & 29.348 &  1.464 & 0.172 & 0.169 \\
        \stwome{}                   & 10.55 &   2.56 &  1.861 &  1.827 &  0.955 & 30.025 & 29.984 &  1.663 & 0.052 & 0.078 \\
    \end{tabular}
    \end{ruledtabular}
\end{sidewaystable}

\clearpage

\section{SI Note: Spin-splitting in 2D perovskites}

To correlate the spin-splitting in 2D perovskites with the aforementioned structural descriptors, we analyze the spin-splitting in the band structure of the perovskites. In this analysis we employ the perovskite geometries optimized using the SCAN XC functional. Analogous to earlier work~\cite{janaStructuralDescriptorEnhanced2021}, we evaluate the spin-splitting of the lowest conduction bands in the $\Gamma$-X, $\Gamma$-Y and $\Gamma$-Z directions using the PBE XC functional including effects of spin-orbit coupling (SOC). Charge densities were converged with an energy convergence criterion of \SI{1E-8}{\eV} on the same $k$-mesh used for the geometry optimizations.

We analyze the spin-splitting of the bands near the $\Gamma$-point through the following functional form~\cite{janaStructuralDescriptorEnhanced2021}
\begin{equation}
    \label{eq:spin_splitting_energies}
    E_{\pm} \left( \mathbf{k} \right) = \frac{\hbar^{2} k^{2}}{2 m} \pm \alphaeff k
\end{equation}
where $E_{+}$ and $E_{-}$ are the energies of the spin-split bands and $\alphaeff$ is the spin-splitting coefficient. This spin-splitting coefficient can be rewritten as
\begin{equation}
    \label{eq:spin_splitting_coefficient}
    \alphaeff = \frac{\deltaEpm}{2 \kzero}
\end{equation}
with $\kzero$ as the characteristic momentum offset of the spin-split bands and $\deltaEpm$ the energy splitting between the bands at $\kzero$ (i.e. $\deltaEpm = E_{+} - E_{-}$). The resulting spin-splitting parameters for the band exhibiting maximum spin-splitting are shown in Table~\ref{tab:spin_splitting}.

\begin{table}[htbp]
    \caption{Spin-splitting parameters of chiral 2D perovskite structures.}
    \label{tab:spin_splitting} 
    \begin{ruledtabular}
    \begin{tabular}{cccc}
        Perovskite          & $\kzero$ ($\SI{}{\per\angstrom}$)         & $\deltaEpm$ (\SI{}{\eV})      & $\alphaeff$ (\SI{}{\eV\angstrom})     \\ \hline
        \smba{}             & 0.026                                     & 0.044                         & 0.841                                 \\
        \ronenea{}          & 0.057                                     & 0.159                         & 1.386                                 \\
        \rtwonea{}          & 0.051                                     & 0.124                         & 1.201                                 \\
        \rfourcl{}          & 0.032                                     & 0.041                         & 0.647                                 \\
        \soneme{}           & 0.002                                     & 0.000                         & 0.076                                 \\
        \stwome{}           & 0.006                                     & 0.002                         & 0.155                                 \\
    \end{tabular}
    \end{ruledtabular}
\end{table}

In Table~\ref{tab:spin_splitting_descriptors} we compare the effective spin-splitting parameter $\alphaeff$ with a range of structural descriptors. The comparison indicates that the structural chirality of the inorganic framework ($\epsMXpar$ and $\epsMXperp$) shows no clear correlation with the spin-splitting parameter. The hydrogen bond asymmetry ($\hbasym$) and in-plane bond angle disparity ($\deltabeta{in}$) show somewhat more correlation with the spin-splitting.

\begin{table}[htbp]
    \caption{Effective spin-splitting parameter against a variety of structural descriptors of chiral 2D perovskite structures.}
    \label{tab:spin_splitting_descriptors} 
    \begin{ruledtabular}
    \begin{tabular}{cccccc}
        Perovskite          & $\alphaeff$ (\SI{}{\eV\angstrom})     & $\epsMXpar$ ($\times$10\textsuperscript{-3})  & $\epsMXperp$ ($\times$10\textsuperscript{-3}) & $\hbasym$ (\SI{}{\angstrom})  & $\deltabeta{in}$ (\SI{}{\degree})     \\ \hline
        \smba{}             & 0.841                                 & +4.534                                        & +1.560                                        & 0.036                         & 7.376                                 \\
        \ronenea{}          & 1.386                                 & +12.629                                       & -9.319                                        & 0.058                         & 18.324                                \\
        \rtwonea{}          & 1.201                                 & -4.394                                        & -2.021                                        & 0.107                         & 15.078                                \\
        \rfourcl{}          & 0.647                                 & +7.762                                        & +0.229                                        & 0.031                         & 10.782                                \\
        \soneme{}           & 0.076                                 & +13.230                                       & -0.648                                        & 0.033                         & 1.026                                 \\
        \stwome{}           & 0.155                                 & -4.302                                        & +0.849                                        & 0.000                         & 1.827                                 \\
    \end{tabular}
    \end{ruledtabular}
\end{table}

\clearpage

\section{SI Note: Machine-learning force fields (MLFFs)}

\subsection{Force field training}

For the investigation of the finite-temperature effects we trained machine-learning force fields (MLFFs) against total total energies, forces and stresses from DFT calculations. The training set was automatically constructed using an on-the-fly active learning scheme from dynamical simulations in an \textit{NpT} ensemble~\cite{jinnouchiOntheflyMachineLearning2019}. Local atomic environments were described using an adaptation of the smooth overlap of atomic positions (SOAP) descriptor~\cite{bartokRepresentingChemicalEnvironments2013}, for which we employed a cutoff for the two-body radial descriptor $\rho_{i}^{\left( 2 \right)}$ of \SI{6.0}{\angstrom}, and a \SI{4.0}{\angstrom} cutoff for the three-body angular descriptor $\rho_{i}^{\left( 3 \right)}$. The atomic positions were broadened using Gaussian distributions with a width of \SI{0.5}{\angstrom}. Both descriptors were expanded on a basis set of spherical Bessel functions and Legendre polynomials, using 8 and 6 Bessel functions for the two-body and three-body descriptors, respectively, with a maximum angular momentum quantum number of $l_{\mathrm{max}} = 2$. The expansion coefficients of this basis set constitute the descriptor for the local atomic environments. To measure the similarity between two local atomic environments, we employed a polynomial kernel function to a power 4, in which the two-body radial and three-body angular descriptor vectors were weighted by 0.1 and 0.9, respectively.

The training of the MLFFs started with a constant temperature simulation at \SI{300}{\K}, using the optimized crystal geometry as the starting point. The initial training run was followed by consecutive constant temperature simulations at \SI{100}{\K} and \SI{450}{\K}, each of which used the final positions and velocities of the previous run as starting point. For all constant temperature simulations the time of the training run was \SI{50}{\ps}. In the final training run we cooled the model systems from \SI{350}{\K} to \SI{50}{\K} over \SI{60}{\ps}, the starting point (i.e. positions and velocities) for these runs was obtained using \SI{10}{\ps} equilibration runs with the respective intermediate force fields. During training the temperature and pressure were controlled using Parinello-Rahman dynamics~\cite{parrinelloCrystalStructurePair1980, parrinelloPolymorphicTransitionsSingle1981} using friction coefficients $\gamma = \SI{5}{\per\ps}$ and $\gamma_{\mathrm{L}} = \SI{5}{\per\ps}$ for the atomic and lattice degrees of freedom, respectively. A time step of $\Delta t = \SI{2}{\fs}$ was used together with an increased mass of the hydrogen atoms of $m_{\ch{H}} = \SI{4}{\amu}$, to enhance the sampling of structures. To limit the size of the force fields, the number of local reference configuration was limited to 2000, which was enforced by allowing configurations to be discarded once saturated. Using the training data of the final MLFFs, we refit the models onto faster descriptors without any Bayesian error estimation to speed up the evaluation for large-scale molecular dynamics production runs. The number of structures in the training sets ($N_{\mathrm{DFT}}$) and the number of local reference configuration for the different elements in the MLFFs ($N_{\mathrm{basis}}$) can be found in Table~\ref{tab:mlff_details}.

\begin{table}[htbp]
    \caption{Size of the training set and number of local reference configurations for various machine-learning force fields (MLFFs) for 2D perovskites.}
    \label{tab:mlff_details} 
    \begin{ruledtabular}
    \begin{tabular}{cccccccc}
         \multirow{2}{*}{XC functional} & \multirow{2}{*}{Training structure}   & \multirow{2}{*}{$N_{\mathrm{DFT}}$ (-)}  & \multicolumn{5}{c}{$N_{\mathrm{basis}}$ (-)}                  \\
                                        &                                       &                                          & \ch{H}    & \ch{C}    & \ch{N}    & \ch{I}    & \ch{Pb}       \\ \hline
         \multirow{3}{*}{SCAN}          & \smba{}                               & 972                                      & 2000      & 2000      & 563       & 1240      & 310           \\
                                        & \pea{}                                & 1030                                     & 2000      & 2000      & 507       & 1175      & 218           \\
                                        & \ba{}                                 & 973                                      & 2000      & 1983      & 471       & 1129      & 226           \\ \hline
         PBE+D3(BJ)                     & \smba{}                               & 933                                      & 2000      & 2000      & 568       & 1190      & 283           \\
    \end{tabular}
    \end{ruledtabular}
\end{table}

\clearpage

\subsection{Force field accuracy}

\begin{figure*}[htbp]
    \includegraphics{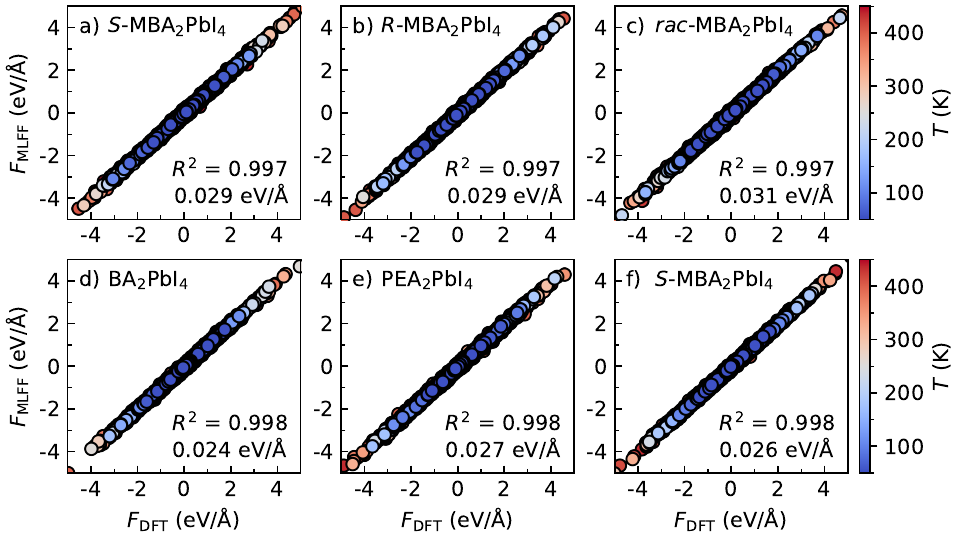}
    \caption{Accuracy of machine-learning force fields (MLFFs) during short heating simulations from \SI{50}{\K} to \SI{450}{\K}. Model performance of MLFF trained on \smba{} structures with the SCAN XC functional for (a) \smba{}, (b) \racmba{} and (c) \rmba{} perovskite. Model performance of (d) \ba{} and (e) \pea{} MLFF trained against SCAN XC functional and (e) \smba{} MLFF trained against PBE+D3(BJ) XC functional. In each subfigure the coefficient of determination $\rsquared$ and mean absolute error $\mae$ in \SI{}{\eV\per\angstrom} between the forces from the MLFF and DFT are given.}
    \label{fig:mlff_model_accuracy} 
\end{figure*}

To check the ability of the MLFFs to reproduce the forces from DFT calculations, we run short simulations in which we heated the systems used during training from \SI{50}{\K} to \SI{450}{\K} during \SI{80}{\ps} using the final MLFFs obtained from training. For these short simulations we decreased the timestep to $\Delta t = \SI{1}{\fs}$ whilst decreasing the hydrogen mass to $m_{\ch{H}} = \SI{1}{\amu}$. The resulting trajectories were divided into 40 equal parts each spanning a temperature range of \SI{10}{\K}. From each of these parts we randomly selected one frame and evaluated it using both the MLFF and DFT calculations, for which we compared the force components of all atoms in Figure~\ref{fig:mlff_model_accuracy}. All models show a high correlation and low error between the MLFF forces and those obtained using DFT.

The model trained for \smba{} shows not only a high accuracy for the \smba{} perovskite structure it was trained against (Figure~\ref{fig:mlff_model_accuracy}a), it can also accurately describe its enantiomer crystal \rmba{} (Figure~\ref{fig:mlff_model_accuracy}b). We attribute this finding to the invariance of the SOAP descriptor to reflections~\cite{bartokRepresentingChemicalEnvironments2013}. More notably, we observe that the MLFF trained on the chiral perovskite can also accurately describe the achiral \racmba{} (Figure~\ref{fig:mlff_model_accuracy}c). Despite small structural differences between the chiral and achiral perovskite structures, the local environments of atoms are similar enough between the two to result in accurate MLFFs, only slightly increasing the $\mae$ for the achiral perovskite. By training MLFFs against perovskites with different ligands (Figure~\ref{fig:mlff_model_accuracy}d and Figure~\ref{fig:mlff_model_accuracy}e) and using a different XC functional (Figure~\ref{fig:mlff_model_accuracy}f) we show the accurate MLFFs can be obtained for a variety of perovskite systems.

\clearpage

\subsection{Molecular dynamics simulations}

\begin{figure*}[htbp]
    \includegraphics{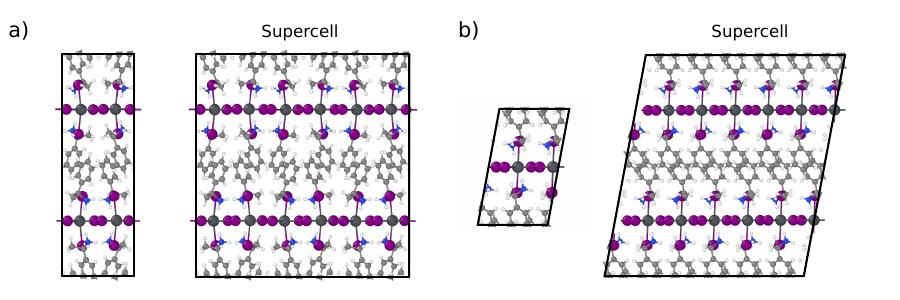}
    \caption{Supercell geometries used in production runs. Unit cell and corresponding supercell for (a) \smba{} and (b) \racmba{}.}
    \label{fig:supercell_geometries} 
\end{figure*}

The production molecular dynamics simulations made use of the same temperature and pressure control settings as used during the MLFF training, but we decreased the simulation time step to $\Delta t = \SI{1}{\fs}$ and set the hydrogen mass to $m_{\ch{H}} = \SI{1}{\amu}$ to sample the system dynamics more accurately, analogous to when we tested the force field accuracy. At each temperature we started five independent constant temperature runs, the total length of each of these runs was \SI{110}{\ps} and the first \SI{10}{\ps} were discarded for system equilibration. The trajectories were sampled every \SI{0.1}{\ps}, resulting in a total of 5000 frames to analyze at each temperature. The structures used during the production runs consist of two inorganic layers and are expanded to similar dimensions in the lateral directions (i.e. a 3 $\times$ 3 supercell in the planar direction). Examples of such structures are shown in Figure~\ref{fig:supercell_geometries} for \smba{} and \racmba{}.

\begin{figure*}[htbp]
    \includegraphics{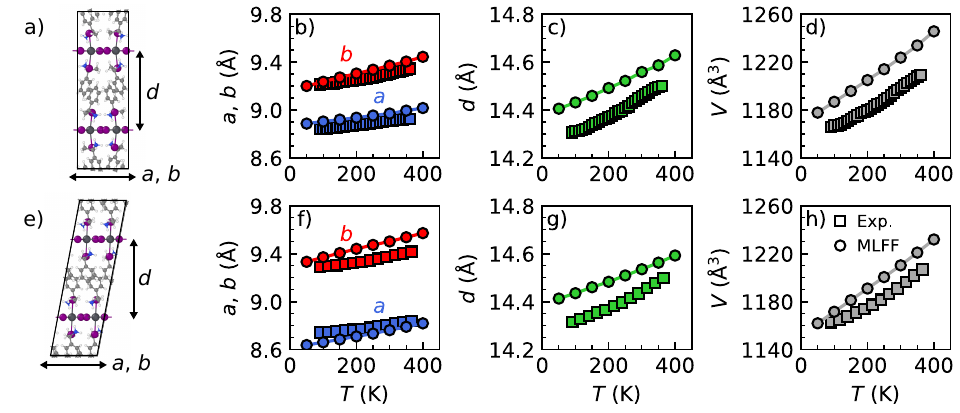}
    \caption{Geometries of 2D perovskite structures from experiments and MLFF simulations. (a) \smba{} perovskite structure and its (b) lateral dimensions, (c) interlayer distances and (d) unit cell volumes. (e) \racmba{} perovskite structure and its (f) lateral dimensions, (g) interlayer distances and (h) unit cell volumes. Experimental data is shown in squares and the simulation data is shown in circles.}
    \label{fig:mlff_geometries} 
\end{figure*}

Focusing on the MLFF trained for \smba{}, we probe its ability to replicate experimentally determined perovskite geometries of \smba{} and \racmba{} perovskites~\cite{dursunTemperatureDependentOpticalStructural2023}. Notably, the single model can accurately model the dimensions of both types of perovskite structures, as is shown in Figure~\ref{fig:mlff_geometries}. The MLFF simulations show a slight overprediction of the volumes of \smba{} (+2.2\%) and \racmba{} (+1.5\%) which appears to stem from the overprediction of the interlayer distances $d$ in the structures.

\clearpage

\section{SI Note: Degree of chirality}

\subsection{Uncertainty quantification of degree of chirality}

\begin{figure*}[htbp]
    \includegraphics{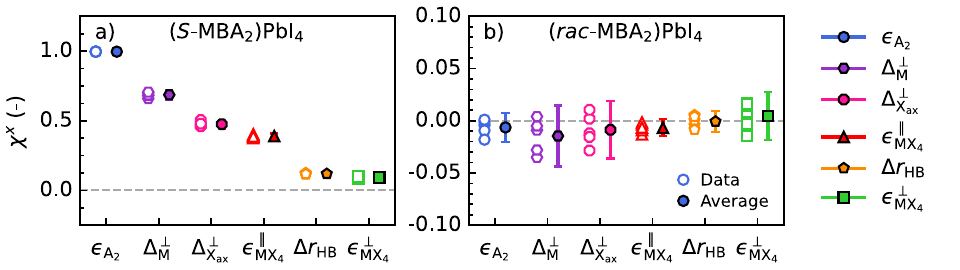}
    \caption{Spread in the values of the structural descriptors determined at \SI{300}{\K}. Degree of chirality for various structural descriptors in (a) \smba{} and (b) \racmba{}. Data points from the five different simulations are shown with the white shapes, the average of the five runs is shown in the colored shapes. All average values have their 95\% confidence interval indicated along with them, when not shown the confidence interval is too narrow to see.}
    \label{fig:chirality_spread} 
\end{figure*}

To quantify the uncertainty in the degree of chirality of various descriptors, we investigate the spread in the values of $\degreechi$ for the five independent runs done at each temperature. As an example, the results of this uncertainty quantification (95\% confidence interval) at \SI{300}{\K} for both \smba{} and \racmba{} are shown in Figure~\ref{fig:chirality_spread}. The figure contains the data points extracted from each of the individual runs (white shapes) and the average and standard of this collection of data points (colored shapes with error bars). The results demonstrate a small spread in the degree of chirality for both perovskites. Notably, for \smba{} the error bars on the data are so small they are within the size of the shapes. For \racmba{} we find that the 95\% confidence intervals contain the degree of chirality of zero. Altogether this indicates the use of the five independent simulation runs is sufficient to identify the difference between chiral and achiral perovskites.

\clearpage

\subsection{Temperature-dependence of chirality in \rmba{}}

\begin{figure*}[htbp]
    \includegraphics{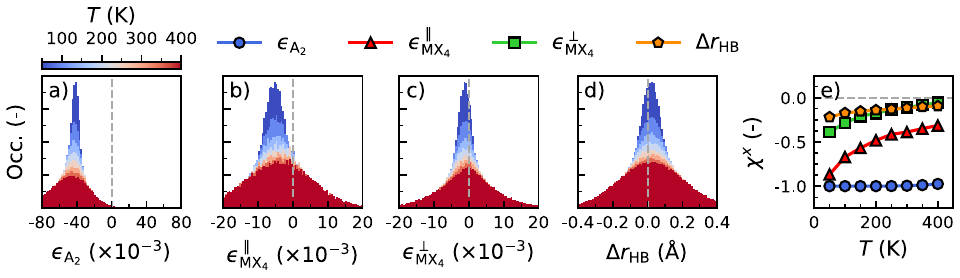}
    \caption{Temperature dependence of the degree of chirality in \rmba{}. Temperature-dependent descriptor distributions for (a) cation chirality, (b) in-plane framework chirality, (c) out-of-plane framework chirality, and (d) hydrogen bond asymmetry. (e) Temperature-dependent degree of chirality.}
    \label{fig:temperature_dependence_rmba} 
\end{figure*}

As done in the main text for \smba{} and \racmba{}, the temperature dependence of the distributions of the structural descriptors and the degree of chirality can be computed for \rmba{}. The result of this analysis is shown in Figure~\ref{fig:temperature_dependence_rmba}. Due to the definition of the degree of chirality $\degreechi$, we find negative values for the various structural descriptors in \rmba{}. We highlight the magnitudes of $\degreechi$ in \rmba{} are the same as for \smba{}, which follows from the mirror symmetry relation between the two enantiomer structures.

\clearpage

\subsection{Temperature-dependence of chirality in achiral perovskites}

\begin{figure*}[htbp]
    \includegraphics{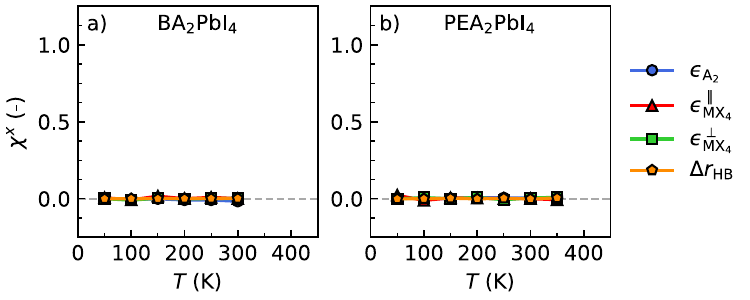}
    \caption{Temperature dependence of the degree of chirality in achiral perovskites. Degree of chirality for (a) \ba{} and (b) \pea{} perovskites.}
    \label{fig:temperature_dependence_achiral} 
\end{figure*}

Analogous to \ch{MBA+}-based perovskites, we can determine the temperature dependence of the degree of chirality in achiral \ba{} and \pea{}. We subject both perovskite structures to a range of temperatures using the earlier-mentioned MLFFs trained for those perovskites. Note, due to structural phase transitions in \ba{} at relatively low temperatures~\cite{biegaDynamicDistortionsQuasi2D2023}, we discarded the initial \SI{60}{\ps} of each individual simulation for \ba{} data to allow for system equilibration. The results of these simulations are shown in Figure~\ref{fig:temperature_dependence_achiral}. For both perovskites, we observe a similar behavior as for achiral \racmba{}, with the structures not showing any structural chirality for any of the investigated descriptors (i.e. $\epsA$, $\epsMXpar$, $\epsMXperp$ and $\hbasym$).

\clearpage

\subsection{Effects of exchange-correlation (XC) functional}

\begin{figure*}[htbp]
    \includegraphics{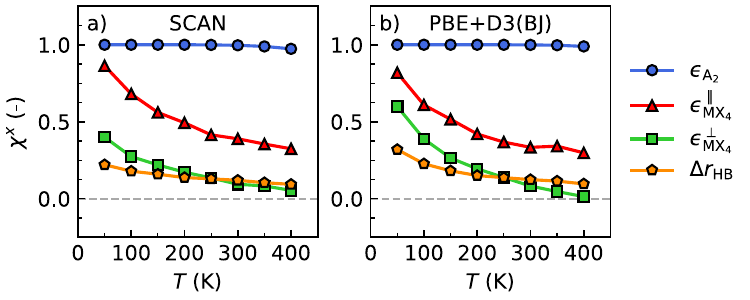}
    \caption{Effects of different exchange-correlation (XC) functionals on the degree of chirality in descriptors. Temperature dependence of $\epsA$, $\epsMXpar$ and $\epsMXperp$ using machine-learning force fields (MLFFs) trained against the (a) SCAN and (b) PBE+D3(BJ) XC functionals.}
    \label{fig:xc_functional_comparison} 
\end{figure*}

All previous molecular dynamics results have been generated using MLFFs trained against data from DFT calculations using the SCAN XC functional. However, since we also trained a MLFF against DFT calculations with the PBE+D3(BJ) (Table~\ref{tab:mlff_details}), we can assess the influence that different XC functionals have on the degree of chirality. We show the temperature dependence of the degree of chirality for both SCAN and PBE+D3(BJ) in Figure~\ref{fig:xc_functional_comparison} and provide some additional data in Table~\ref{tab:xc_functional_comparison}. Both functionals exhibit a decrease of the degree of chirality for the inorganic framework ($\epsMXpar$ and $\epsMXperp$), but not for the arrangement of the organic cations ($\epsA$). At high temperatures (i.e. \SI{400}{\K}), irrespective of the functional the MLFF was trained against, a similar degree of chirality is found for all three descriptors for both MLFFs. This indicates that both models result in similar dynamics at such elevated temperatures. In contrast, some differences in the degree of chirality are found at low temperatures (i.e. \SI{50}{\K}), where PBE+D3(BJ)-based MLFF predicts a slightly lower degree of chirality for the in-plane frame work chirality ($\epsMXpar$), but a somewhat larger degree of chirality for the out-of-plane framework chirality ($\epsMXperp$). We hypothesize these differences stem from the explicit inclusion of van der Waals forces in the PBE+D3(BJ) functional, which results in stronger interactions between the organic cations and inorganic framework, in turn leading to more severe out-of-plane distortions of the inorganic framework. Based on the more accurate unit cell volumes (SI Note~\ref{section:density_functional_theory}) and previous benchmarks of XC functionals for hybrid halide perovskites~\cite{bokdamAssessingDensityFunctionals2017, lahnsteinerFinitetemperatureStructureMAPbI32018}, we deem the dynamics from the SCAN-based MLFF to be more accurate.

\begin{table}[htbp]
    \caption{Degrees of chirality for different XC functionals at different temperatures.}
    \label{tab:xc_functional_comparison} 
    \begin{ruledtabular}
    \begin{tabular}{ccccc}
        $T$ (K)                         & XC functional         & $\chi^{\epsA}$ (-)        & $\chi^{\epsMXpar}$ (-)    & $\chi^{\epsMXperp}$ (-)   \\ \hline
        \multirow{2}{*}{50}             & SCAN                  & 1.000                     & 0.864                     & 0.399                     \\
                                        & PBE+D3(BJ)            & 1.000                     & 0.818                     & 0.599                     \\ \hline
        \multirow{2}{*}{400}            & SCAN                  & 0.973                     & 0.324                     & 0.054                     \\
                                        & PBE+D3(BJ)            & 0.989                     & 0.298                     & 0.013                     \\
    \end{tabular}
    \end{ruledtabular}
\end{table}

\clearpage

\section{SI Note: Chirality transfer}

\subsection{Effects of exchange-correlation (XC) functional}

\begin{figure*}[htbp]
    \includegraphics{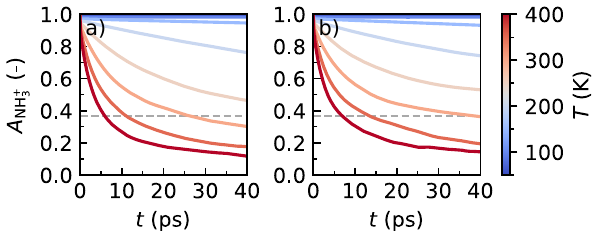}
    \caption{Dependency of cation headgroup orientational autocorrelation function $A_{\ch{NH3+}}$ on exchange-correlation (XC) functional for the (a) SCAN and (b) PBE+D3(BJ) XC functionals. The typical rotation time for the cation headgroups in \smba{} at \SI{400}{\K} is \SI{6.0}{\ps} and \SI{6.9}{\ps} for the MLFFs trained against SCAN and PBE+D3(BJ), respectively.}
    \label{fig:nh3_headgroup_xc_dependence} 
\end{figure*}

To assess the influence of the XC functional on the dynamics of the \ch{NH3+} group, we compare the orientational autocorrelation functions in Figure~\ref{fig:nh3_headgroup_xc_dependence}. The simulations with both the SCAN (Figure~\ref{fig:nh3_headgroup_xc_dependence}a) and PBE+D3(BJ) (Figure~\ref{fig:nh3_headgroup_xc_dependence}b) functionals, exhibit a very similar time and temperature dependence. Interestingly, the typical rotation time of the \ch{NH3+} group $\tau_{\ch{NH3+}}$ for both functionals at \SI{400}{\K} is similar, at \SI{6.0}{\ps} and \SI{6.9}{\ps} for the MLFF trained using SCAN and PBE+D3(BJ), respectively. We connect this similarity in the dynamics at elevated temperatures to the above-mentioned similarity in the degree of chirality between the two MLFFs.

\clearpage

\subsection{Cation reorientations}

\begin{figure*}[htbp]
    \includegraphics{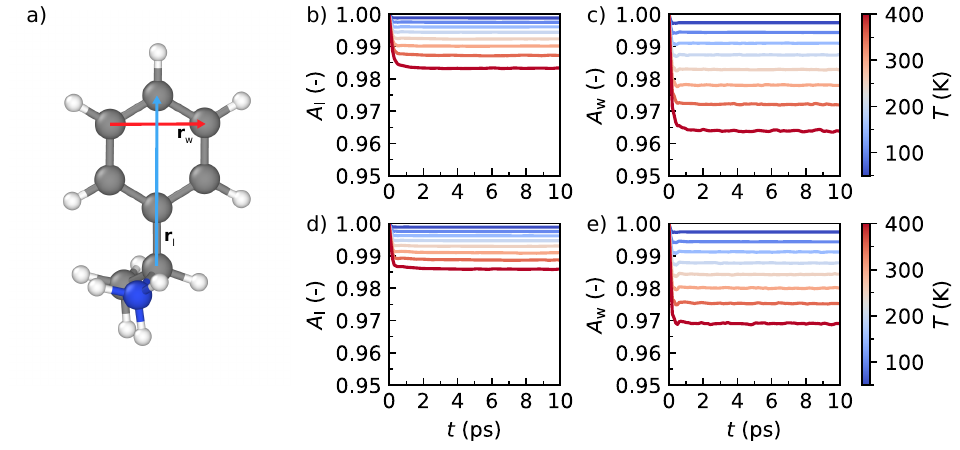}
    \caption{Orientational correlation function for organic cations. (a) The length $\mathbf{r}_{\mathrm{l}}$ and width $\mathbf{r}_{\mathrm{w}}$ vectors used to define the cation orientation. The orientational autocorrelation of the (b) length and (c) width vectors in \smba{} and (d) length and (e) width vectors in \racmba{}.}
    \label{fig:cation_orientation} 
\end{figure*}

In addition to determining the orientational autocorrelation of the $\ch{N}-\ch{H}$ bonds in the \ch{NH3+} groups, we can also employ the autocorrelation function to assess the persistence of the orientation of the organic cations. To do so, we use orientation vectors that span the length ($\mathbf{r}_{l}$) and width ($\mathbf{r}_{w}$) of the \ch{MBA+} cations, as shown in Figure~\ref{fig:cation_orientation}a. The vectors can be defined using the atomic fingerprints from SI Note~\ref{section:atomic_fingerprints} as shown in Table~\ref{tab:mba_orientation_vectors_fingerprints}. The orientational autocorrelation functions of these vectors (Figure~\ref{fig:cation_orientation}b-e) level off at values close to 1. This indicates a highly persistent orientation of the \ch{MBA+} cations in both \smba{} (Figure~\ref{fig:cation_orientation}b-c) and \racmba{} (Figure~\ref{fig:cation_orientation}d-e), where the organic cations do not exhibit any rotations.

\begin{table}[htbp]
    \caption{Atom fingerprints used to determine the orientation of the \ch{MBA+} cation.}
    \label{tab:mba_orientation_vectors_fingerprints} 
    \begin{ruledtabular}
    \begin{tabular}{ccccc}
        Vector                  & Element 1     & Atom fingerprint 1        & Element 2     & Atom fingerprint 2        \\ \hline
        $\mathbf{r}_{l}$        & \ch{C}        & \texttt{H32C16N1}         & \ch{C}        & \texttt{H52C18N5}         \\
        $\mathbf{r}_{w}$        & \ch{C}        & \texttt{H46C16N4}         & \ch{C}        & \texttt{H46C16N4}         \\
    \end{tabular}
    \end{ruledtabular}
\end{table}

\clearpage

\section{SI Note: Distortions and hydrogen bonds in perovskites}

To compare the degree with which perovskite structures are distorted, we compare the values of several structural descriptors for 2D perovskite structures with the \ch{MBA+}, 1\ch{NEA+} and 2\ch{NEA+} cation. The results of this analysis are shown in Table~\ref{tab:distortion_comparison}. Distortions of the inorganic framework are captured by the bond angle variance $\sigmatwo$ and the bond length variance $\deltad$. For all investigated cations, the chiral structures (i.e. \textit{S}-/\textit{R}-) have larger distortions in the inorganic framework than the achiral racemic counterparts, which is in line with earlier work in literature~\cite{janaOrganictoinorganicStructuralChirality2020, sonUnravelingChiralityTransfer2023}. We focus ourselves on the nature of the hydrogen bonds between the organic cations and inorganic framework in the various perovskites, by investigating the hydrogen bond lengths in the inorganic cages. To do so, we report the hydrogen bond length of the strongly ($\bar{r}^{\mathrm{strong}}_{\mathrm{HB}}$) and weakly ($\bar{r}^{\mathrm{weak}}_{\mathrm{HB}}$) bound cation to the cages as well as the average hydrogen bond length ($\Bar{r}_{\mathrm{HB}}$) of these cations. Here we find that for all cations the chiral structures (\textit{S}-/\textit{R}-) bind the cations more weakly, as illustrated by the longer hydrogen bond lengths, when compared to the racemic perovskite structures. We hypothesize the larger framework distortions and more weakly bound organic cations come about due to the breaking of symmetry that results from the chiral cations, which also distorts the framework and weakens the hydrogen bonds.

\begin{table}[htbp]
    \caption{Information of inorganic framework distortions and hydrogen bond strength for various 2D perovskite structures with chiral organic cations. The data is from structures optimized using the SCAN XC functional.}
    \label{tab:distortion_comparison} 
    \begin{ruledtabular}
    \begin{tabular}{cccccc}
        Structure           & $\sigma^{2}$ ($^{\circ 2}$)   & $\Delta d$ ($\times 10^{-5}$)     & $\bar{r}^{\mathrm{strong}}_{\mathrm{HB}}$ (\SI{}{\angstrom})  & $\bar{r}^{\mathrm{weak}}_{\mathrm{HB}}$ (\SI{}{\angstrom})    & $\Bar{r}_{\mathrm{HB}}$ (\SI{}{\angstrom})    \\ \hline
        \smba{}             & 20.17                         & 29.38                             & 2.605                                                         & 2.640                                                         & 2.623                                         \\
        \rmba{}             & 20.17                         & 29.38                             & 2.605                                                         & 2.640                                                         & 2.623                                         \\
        \racmba{}           & 14.04                         & 18.19                             & 2.595                                                         & 2.595                                                         & 2.595                                         \\ \hline
        \sonenea{}          & 42.48                         & 75.48                             & 2.364                                                         & 2.408                                                         & 2.386                                         \\
        \ronenea{}          & 43.21                         & 70.85                             & 2.358                                                         & 2.416                                                         & 2.387                                         \\
        \raconenea{}        & 22.58                         & 20.88                             & 2.345                                                         & 2.345                                                         & 2.345                                         \\ \hline
        \stwonea{}          & 48.15                         & 47.24                             & 2.325                                                         & 2.423                                                         & 2.374                                         \\
        \rtwonea{}          & 49.53                         & 48.14                             & 2.319                                                         & 2.427                                                         & 2.373                                         \\
        \ractwonea{}        & 23.43                         &  8.97                             & 2.315                                                         & 2.315                                                         & 2.315                                         \\ 
    \end{tabular}
    \end{ruledtabular}
\end{table}

\clearpage

\bibliography{supporting}